\documentclass[3p,sort&compress]{elsarticle}

\usepackage{latexsym}
\usepackage{epsfig}
\usepackage{amsmath}
\usepackage{amsfonts}
\usepackage{amssymb}
\usepackage{graphicx}
\usepackage{pgf}
\usepackage{enumerate}
\usepackage{amssymb}
\usepackage{caption}
\usepackage{subcaption}
\usepackage{lineno}
\modulolinenumbers[5]
\usepackage{xcolor}
\usepackage{hyperref}
\usepackage{float}

\usepackage{algorithm}
\usepackage[noend]{algpseudocode}

\graphicspath{{figures/}}

\makeatletter
\def\BState{\State\hskip-\ALG@thistlm}
\makeatother

\journal{J.~non-Newtonian Fluid Mech.}

\begin{document}

\begin{frontmatter}

\title{Squeeze cementing of micro-annuli: a visco-plastic invasion flow}

\author[UBC]{Mahdi Izadi}

\author[STRATH]{Emad Chaparian
\cortext[correspondingauthor]{Corresponding author: emad.chaparian@strath.ac.uk (Emad Chaparian)}}

\author[UBC]{Elizabeth Trudel}

\author[UBC]{Ian Frigaard}

\address[UBC]{Department of Mechanical Engineering \& Mathematics, University of British Columbia, Vancouver, BC, Canada}
\address[STRATH]{James Weir Fluid Laboratory, Department of Mechanical \& Aerospace Engineering, University of Strathclyde, Glasgow, United Kingdom}


\begin{abstract}
Squeeze cementing is a process used to repair leaking oil and gas wells, in which a cement slurry is driven under pressure to fill an uneven leakage channel. This results in a Hele-Shaw type flow problem involving a yield stress fluid. We solve the flow problem using an augmented Lagrangian approach and advect forward the fluid concentrations until the flow stops.  A planar invasion and a radial (perforation hole) invasion flow are studied. The characteristics of the flow penetration are linked to the channel thickness profile. The distribution of streamlines, flowing and non-flowing zones, evolves during the invasion flow. An interesting aspect of the results is the extreme variability in penetration metrics computed. These depend not only on the stochastic nature of the microannulus thickness, which has significant natural variation in both azimuthal and axial directions, but also on the ``luck'' of where the perforation hole is, relative to the larger/smaller microannulus gaps. This may explain the unreliability of the process.
\end{abstract}

\begin{keyword}
Hele-Shaw flows; yield-stress fluid; squeeze cementing
\end{keyword}

\end{frontmatter}


\section{Introduction}
\label{sec:intro}

This paper concerns the modelling of the invasion of a viscoplastic fluid (a microfine cement slurry) into an uneven narrow channel (microannulus). This \emph{squeeze} cementing operation occurs in the repair of leaking oil and gas wells, which may either be emitting greenhouse gases (GHG) or allowing oil seepage to surface, both flows occurring along the channel to be filled.

When oil and gas wells are constructed, a critical part of the completion process is an operation \emph{primary} cementing, whereby cement is pumped into the narrow annular space between the borehole and a steel casing (through which the produced hydrocarbons will flow); see Nelson and Guillot \cite{Nelson2006}. This cement sheath is vital for mechanical support, to protect the steel from corrosive formation brine and to seal hydraulically by adhering to both the steel casing and the borehole wall, referred to as well integrity. In a typical well, a number of casings will be cemented concentrically inside one another as the well extends deeper.

\begin{figure}[!h]
	\centering
	\includegraphics[width=0.85\linewidth]{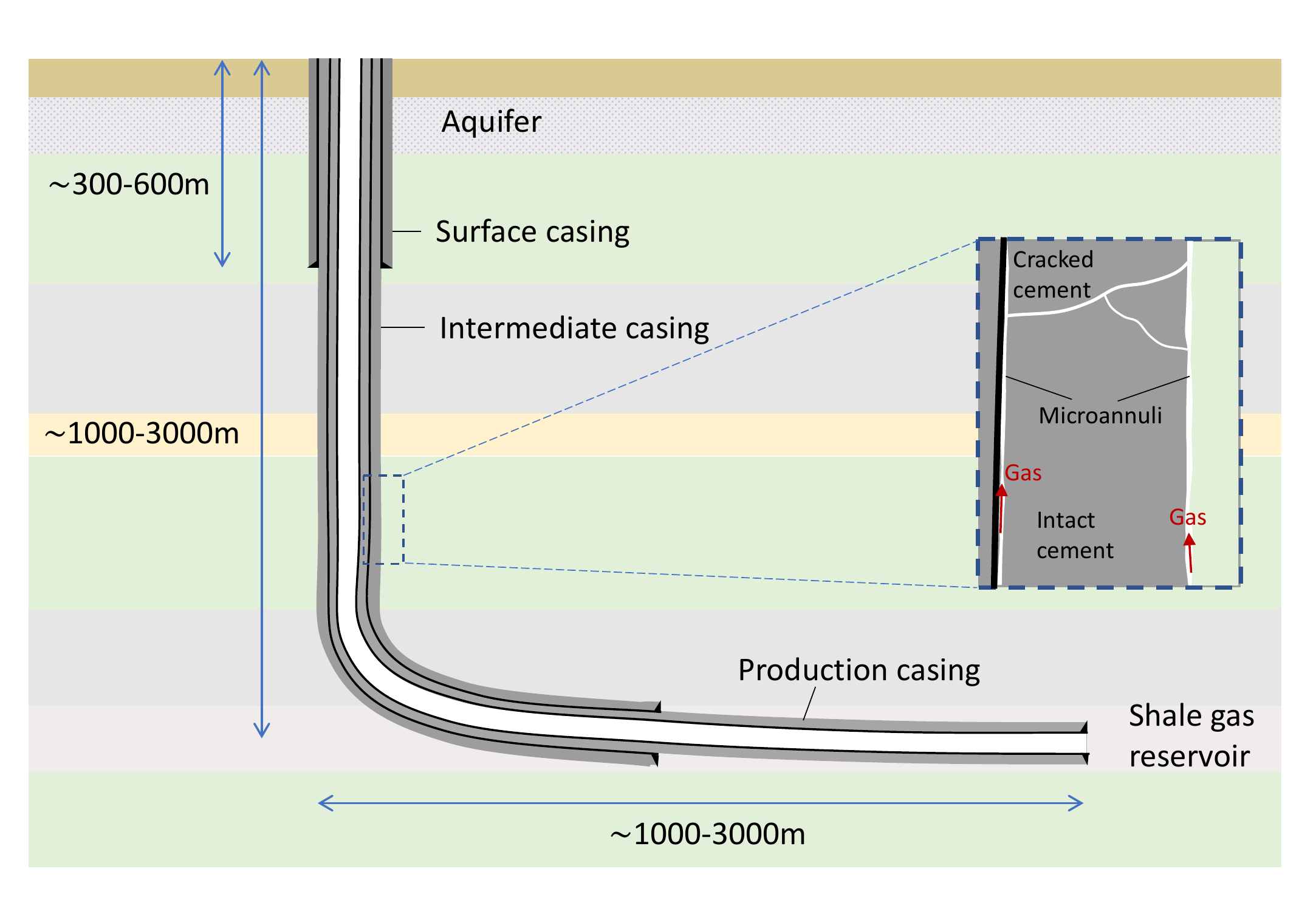}	
	\caption{Schematic of typical unconventional gas wells in British Columbia, Canada. Inset shows microannuli and cracked cement, compromising well integrity.}
	\label{fig:schematic1}
\end{figure}

Wellbore leakage occurs fairly frequently and is an indication of failure to provide a hydraulic seal around the cemented annulus. A recent study found that 13.9 \% of all wells in British Columbia (BC), Canada, registered an instance of surface casing vent flow (SCVF) at some point throughout their operating life, meaning that leaking fluids are released through the surface casing assembly. For wells drilled between 2010 and 2018, 28.5 \% reported an instance of SCVF \cite{Trudel2019}. Leakage pathways associated with SCVF include microannuli, cracks within the primary cement sheath, debonding between the annular cement sheath and the interface, and in rare cases, the cement sheath itself may provide a pathway to leakage if its permeability has been compromised during the primary cementing operation (e.g. from excessive contamination); see Fig.~\ref{fig:schematic1}. It is widely acknowledged that some form of coherent microannular gap is needed for significant leakage to surface, along the casing. The microannulus may arise from cement shrinkage or might result from debonding following pressurization of the well casing (e.g.~during a hydraulic fracturing operation). For our purposes, we simply assume that such a gap occurs.

Microannuli and their role in well leakage has become of interest recently not only for GHG emission concerns, but also as part of the design process for potential leakage of $CO_{2}$ in carbon capture and storage (CCS) operations. Thus, from the CCS direction there are a number of simple models that assign an effective permeability to a well, to represent the microannulus leakage \cite{Doherty2017,Lackey2019,White2020}. In well decommissioning on the other hand, the microannulus is generally assigned a thickness and interpreted as a uniform narrow gap. With a view to developing probabilistic approaches to assessing leakage risk, researchers have developed models in which cement permeability, microannulus thickness and crack dimensions are all considered uncertain inputs and sampled from a distribution
\cite{Ford2017,Willis2019,Johnson2021,AlRamadan2021}. Uniform microannulus thicknesses of $5 - 300 \mu m$ have been used.

In addition to the works discussed above,  a number of experimental studies have attempted to measure either flow through a microannulus, to determine its effective permeability, or microannulus size/geometry directly. These experiments are typically conducted on scaled-down sections of wells and aim to recreate downhole conditions such as temperature and pressure cycling. Computer tomography (CT) has been used to determine the size of leakage pathways \cite{Vralstad2015,Skorpa2018,Vralstad2020,Ogienagbon2021}, showing that microannulus geometries are highly variable in nature. A complex network of pathways with non-uniform thickness were observed in their samples with connectivity of pathways in the axial direction. However, the technique is  limited in regards to microannulus size by the resolution of the CT scanner, as discussed in Vr{\aa}lstad, Skorpa and co-authors \cite{Vralstad2015,Skorpa2018, De2016}. Similiar non-destructive visualization of microannuli was conducted by Yang \emph{et al.}~\cite{Yang2021}, with improved resolution. Both micro-CT and electron microscopy (ESEM) were used and resolutions of $11.92 \mu m$ for the micro-CT scanner and $0.05 \mu m$ with ESEM were achieved. This study determined the presence of leakage pathways at the cement/formation interface and a varied gap size according to formation properties. Garcia Fernandez \emph{et al.}~\cite{GarciaFernandez2019} recreated microannuli experimentally under various conditions and characterised the distribution of thickness values (as right skewed distributions such as gamma and lognormal distributions). Thicknesses with range from $0$ to $50 \mu m$ were found for thermally debonded annuli and $0$ to $1000 \mu m$ for larger microannuli, where a residual layer was found between the cement and the casing/formation. Their results showed highly varied microannulus aperture size in the azimuthal direction. This conclusion is also supported by Stormont \emph{et al.}~\cite{Stormont2018}, who observed hydraulic apertures in the range of $0$ to $118 \mu m$. Results from experimental studies thus point to microannuli being highly varied in the azimuthal direction but presenting some continuity in the well axis direction and with microannulus thicknesses ranging from $0$ to upwards of $200 \mu m$. Taking these basic features Trudel and Frigaard \cite{Trudel2023} derived a probabilistic model for microannulus thickness along sections of well. The base model parameters were calibrated against the distribution of well leakage data collected from over 3000 wells. Later, we use the same model to generate representative microannuli for our study.

\begin{figure}[!h]
	\centering
	\includegraphics[width=0.95\linewidth]{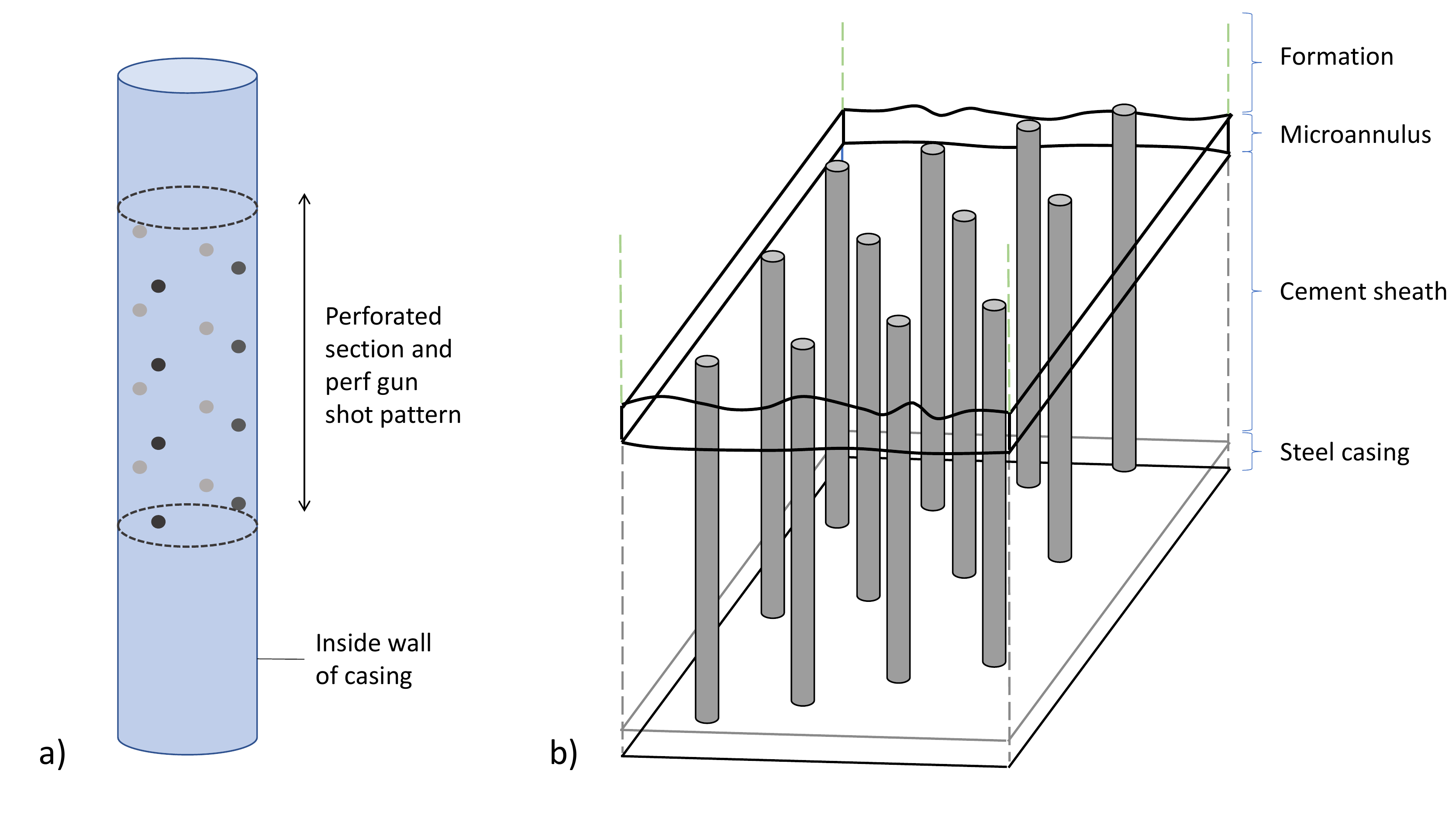}	
	\caption{a) Schematic of typical perforation gun pattern on the inside of the casing. b) Unwrapped irregular microannulus with perforation holes through casing and into formation.}
	\label{fig:schematic2}
\end{figure}

\begin{figure}[!h]
	\centering
	\includegraphics[width=0.95\linewidth]{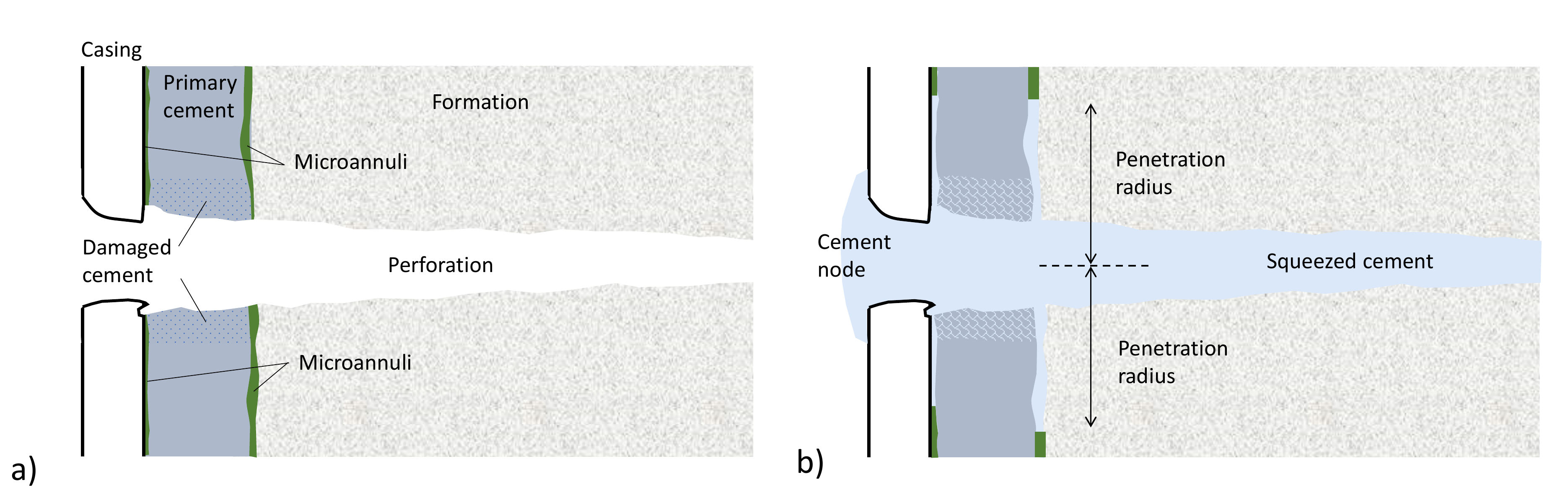}	
	\caption{Schematic of squeeze cementing of a single perforation: a) directly after perforation; b) at end of squeeze cementing, indicating penetration radius.}
	\label{fig:schematic3}
\end{figure}

This paper targets the repair of microannular well leakage, through an operation called squeeze cementing. For a leaking well, having identified the leaking zone, the steel casing and cement sheath are perforated using shaped charges; Fig.~\ref{fig:schematic2}. After some washing of debris from the holes, a section of the well is isolated longitudinally and a cement slurry is pumped into the well under pressure. The cement enters into the (large diameter) perforations and any induced fractures/crevices in the surrounding formation, but the pump pressure is generally kept low enough so as not to hydraulically fracture. In the larger channels the cement slurry slows and begins to pack, back towards the well. The arrest of the slurry happens via a combination of rheological and filtration mechanisms, not fully understood \cite{Izadi2021}. As the resistance downstream builds, the cement slurry is also forced laterally along the well into the narrower cavities of the microannuli and any damaged zones near the perforation; see Fig.~\ref{fig:schematic3}.
It is of interest to understand this invasion process and eventually be able to predict features such as the invasion depth of the slurry, as a function of modelled geometry, at least in a probabilistic sense. This explains the main motivation and direction of this paper.

Whereas much of the squeeze cementing flow will occur in the near wellbore region, from the leakage perspective the microannulus flow is of key interest. This consists of invasion of a viscoplastic slurry into a narrow uneven channel. Squeeze cementing shares features with at least two other industrial processes. In terms of the geometric uncertainty away from the wellbore, grouting processes are analogous \cite{Stille2015,ElTani2017}. Regarding the microannulus flow, this is quite similar to the initial primary cementing operation performed on the well, which is typically modelled as a Hele-Shaw flow along a narrow eccentric annulus \cite{Bittleston2002,Maleki2017}. There are three significant differences. First, the invasion of the cement into the microannulus is not displacing a rheologically complex fluid, such as the drilling mud in primary cementing. Typically, water is used to clean the perforations from debris in squeeze cementing and may therefore be in the microannulus, but no other preflush is used. Secondly, the geometry in primary cementing is a uniform eccentric annulus whereas the microannulus geometry is likely more irregular. Thirdly, given the median size of cement particles and typical size of microannular gap, it is clear that for squeeze cementing we are often close to the limit where the cement slurry may be regarded as a single phase continuum, as opposed to a suspension of particles, as  \cite{Izadi2021}. However, depending on operator and service company involved, the use of microfine cements is generally recommended, which have particle sizes $\sim 5-20 \mu m$.

Flows of a viscoplastic fluid past uneven cavities \cite{Roustaei2015a,Roustaei2015b}, and along uneven fracture geometries \cite{Roustaei2016}, have been studied in depth. Hele-Shaw approaches remain valid in the limits of long-thin geometries. More recently, there have been a number of studies of displacement flows through uneven wellbore geometries \cite{Renteria2019,Skadsem2019a,Skadsem2019b,Etrati2020,Renteria2021,Espinoza2021,Sarmadi2022,Varges2022}, which have used both Hele-Shaw and Navier-Stokes formulations.

An outline of the paper is as follows.

\section{Model derivation}
\label{sec:model}

In this paper we look at the pressure-driven invasion of a visco-plastic (cement) slurry into an uneven narrow cavity (the microannulus), which is itself filled with another miscible fluid (e.g.~water). In general, the setup involves a micro-annulus domain that is bounded by surfaces of at least 2 types: injection/invasion surface(s) and far-field/outflow surfaces. The typical micro-annulus gap has a length-scale $2\hat{H}_0 \sim 100~\mu$m and the lateral dimensions of the micro-annulus domain are $O(\hat{L}) \sim 1$m. The disparity of length-scales ($\delta = \hat{H}_0 /\hat{L} \sim 10^{-4} \ll 1$), suggests that curvature may be neglected and we may effectively unwrap the micro-annulus domain into a two-dimensional domain with lateral dimensions $(\hat{x},\hat{y})$ and transverse coordinate $\hat{z} \in [-\hat{H}(\hat{x},\hat{y}),\hat{H}(\hat{x},\hat{y})]$.

In the above, as discussed in \S \ref{sec:intro}, the situation is rather similar to the primary cementing flows of \cite{Bittleston2002,Maleki2017} in terms of underlying model. However, primary cementing flows are strongly influenced by buoyancy effects, which become significant/evident when the buoyancy number $b$:
\begin{equation}\label{eq:buoyancy}
  b = \frac{\Delta \hat{\rho} \hat{g} \hat{H}_0^2}{\hat{\mu}_s \hat{V}_0} \gtrsim 10.
\end{equation}
Here $\Delta \hat{\rho}$ represents a density difference between fluids, $\hat{g}$ is the gravitational acceleration, $\hat{\mu}_s$ and $\hat{V}_0$ are representative viscosities and velocities of the slurry, respectively. In general for squeeze cementing, due to the reduced $\hat{H}_0$ compared to primary cementing, we will have $b \lesssim 1$. While there may still be extreme situations in which buoyancy, between the cement and in-situ fluid plays a role, we initially ignore. This also simplifies significantly as the direction of gravity, relative to $(\hat{x},\hat{y})$, would otherwise change significantly around the annulus.

Under the above assumptions regarding buoyancy, together with $\delta \ll 1$ and $\delta Re \ll 1$, the leading order\footnote{Here $Re$ is a Reynolds number based on $\hat{H}_0$, $\hat{\mu}_s$, $\hat{V}_0$ and a representative density. We will not use it further.} terms in the Navier-Stokes equations are as follows:
\begin{equation}
\label{eq:momentum}
\left.
\begin{array}{l}
0 = -\displaystyle \frac{\partial \hat{p}}{\partial \hat{x}} + \frac{\partial \hat{\tau}_{xz} }{\partial \hat{z}}, \\[10pt]
0 = -\displaystyle \frac{\partial \hat{p}}{\partial \hat{y}} + \frac{\partial \hat{\tau}_{yz} }{\partial \hat{z}}, \\[10pt]
0=-\displaystyle \frac{\partial \hat{p}}{\partial \hat{z}}
\end{array} \right\} ~~~~ \Rightarrow ~~~~
\left\{
\begin{array}{l}
\hat{\tau}_{xz} = \hat{z} \displaystyle \frac{\partial \hat{p}}{\partial \hat{x}} , \\[10pt]
\hat{\tau}_{yz} = \hat{z} \displaystyle \frac{\partial \hat{p}}{\partial \hat{y}} , \\[10pt]
\hat{p} = \hat{p}(\hat{x},\hat{y})
\end{array} \right.
\end{equation}
\begin{equation}\label{eq:incomp}
0 = \hat{\nabla} \cdot \hat{\mathbf{u}}.
\end{equation}
The constitutive laws are also simplified, being based on the leading order shear flow components of the deviatoric stress and strain rate tensors:
\begin{equation}\label{eq:const}
\left\{
\begin{array}{ll}
\hat{\boldsymbol{\tau}} = \left( \hat{\kappa} \Vert \hat{\dot{\boldsymbol{\gamma}}} \Vert^{n-1} + \displaystyle{\frac{\hat{\tau}_y}{\Vert \hat{\dot{\boldsymbol{\gamma}}} \Vert}} \right) \hat{\dot{\boldsymbol{\gamma}}} & \mbox{iff}\quad \Vert \hat{\boldsymbol{\tau}} \Vert > \hat{\tau}_y, \\[2pt]
\hat{\dot{\boldsymbol{\gamma}}} = 0 & \mbox{iff}\quad \Vert \hat{\boldsymbol{\tau}} \Vert \leqslant \hat{\tau}_y,
\end{array} \right.
\end{equation}
\begin{equation}
\Vert \hat{\dot{\gamma}} \Vert = \sqrt{ \left( \frac{\partial \hat{u}}{\partial \hat{z}} \right)^2 + \left( \frac{\partial \hat{v}}{\partial \hat{z}} \right)^2 }.
\end{equation}
The parameters above are the yield stress ($\hat{\tau}_y$), consistency ($\hat{\kappa}$), and power law index ($n$), in the usual Herschel-Bulkley law.

The micro-annulus might initially be filled with gas, brine or oil, depending on the type of leakage. Typically, a water-based preflush is pumped ahead of the slurry to wash residual fluids out as well as debris from the perforation. In specialised cases a more rheologically complex preflush may be used \cite{Lichinga2020}. Our assumption is that whatever the in situ fluid is, it is of much smaller viscosity than the invading cement slurry. Thus, although we have a Hele-Shaw type problem we do not expect fingering type instabilities. Due to the large viscosity ratio we assume that the fluid distribution across the gap is uniform, i.e.~non-dispersive. We model the cement-preflush flow as a miscible displacement, characterised by the volume fraction $c(\hat{x},\hat{y},\hat{t})$ of the slurry.
\begin{equation}\label{eq:conceqn}
\frac{\partial c}{\partial \hat{t}} +  \hat{\mathbf{\nabla}} \cdot [ c\hat{\mathbf{u}} ] = \hat{\mathbf{\nabla}} \cdot [\hat{D}_d \hat{\nabla} c].
\end{equation}
Here $\hat{D}_d$ represents the effects of diffusive transport which can be a combination of molecular diffusion and dispersion effects. We will assume that $\hat{D}_d/(\hat{V}_0 \hat{L}) \ll 1$ (large P\'{e}clet number limit), so that the main transport method is advective. On averaging across the micro-annulus gap, and neglecting the diffusive terms, the gap-averaged concentration $\bar{c}$ satisfies the simplified equation:
\begin{equation}\label{eq:conceqn0}
\frac{\partial}{\partial \hat{t}}[ \hat{H} \bar{c} ] +  \hat{\mathbf{\nabla}} \cdot [ \hat{H} \bar{c}(\hat{\bar{u}}, \hat{\bar{v}} ) ] = 0,
\end{equation}
where now $\hat{\mathbf{\nabla}}$ operates only in the $(\hat{x},\hat{y})$-plane.

The concentration is used, together with the properties of both fluids, to define \emph{mixture} rheological properties.
The mixture laws we use are based on simple linear interpolation of stresses, using the slurry properties as the main scale as this provides the most resistance to the pressure applied. To clarify, a velocity scale $\hat{V}_0$ is based on a viscous balance, i.e.
\[ \frac{\hat{\kappa}_s \hat{V}_0^{n_s} }{\hat{H}_0^{n_s}} = \frac{\hat{p}_0}{\hat{L}}\hat{H}_0 , \]
with $\hat{\kappa}_s$ and $n_s$ the slurry consistency and power law index. The viscosity scale $\hat{\mu}_s$ used above is $\hat{\mu}_s = \hat{\kappa}_s \hat{V}_0^{n_s-1}/\hat{H}_0^{n_s-1}$. The mixture power law index is:
$n = c n_s + (1-c) n_p$, with the $p$ subscript denoting preflush properties. The mixture yield stress and consistency are defined as:
\begin{equation}\label{eq:mixture}
  \hat{\tau}_y = c \hat{\tau}_{y,s} + (1-c) \hat{\tau}_{y,p}, ~~~~~~ \hat{\kappa} \frac{\hat{V}_0^{n} }{\hat{H}_0^{n}} = c \hat{\kappa}_s \frac{\hat{V}_0^{n_s} }{\hat{H}_0^{n_s}} + (1-c) \hat{\kappa}_p \frac{\hat{V}_0^{n_p} }{\hat{H}_0^{n_p}}.
\end{equation}
Often the preflush will be Newtonian, in which case $\hat{\tau}_{y,p} = 0$, $n_p=1$ and $\hat{\kappa}_p$ is the viscosity. Of course, (\ref{eq:mixture}) is one of many possible mixture laws and there is no particular justification. For the fluid pairs we shall consider, with much less viscous preflush, we effectively assume that
\[ \hat{\kappa}_s \frac{\hat{V}_0^{n_s} }{\hat{H}_0^{n_s}} \gg \hat{\kappa}_p \frac{\hat{V}_0^{n_p} }{\hat{H}_0^{n_p}} ~~~~\mbox{and}~~~~\hat{\tau}_{y,s} \gg \hat{\tau}_{y,p}  . \]

\subsection{Stream-function formulation}

From the above reduced model, we can follow the usual Hele-Shaw derivation for a visco-plastic fluid \citep{Bittleston2002,Maleki2017}, given briefly here. First, we see that the areal flow rates are divergence-free in the $(\hat{x},\hat{y})$-plane, which leads to definition of the stream function $\hat{\psi}$:
\begin{equation}\label{eq:streamfunction}
  \left( \frac{\partial \hat{\psi}}{\partial \hat{y}}, -\frac{\partial \hat{\psi}}{\partial \hat{x}} \right) = \int_0^{\hat{H}} \left( \hat{u} , \hat{v} \right) ~\text{d} \hat{z} = \hat{H} \left( \hat{\bar{u}} , \hat{\bar{v}} \right) .
\end{equation}
Here we have assumed symmetry of the flow about $\hat{z} = 0$.
From (\ref{eq:momentum}) and (\ref{eq:const}) it follows upon rearranging and integrating 2 times, that $ \left( \hat{\bar{u}} , \hat{\bar{v}} \right)$ is parallel to the pressure gradient $\hat{\nabla} \hat{p}$, flowing down the gradient.

Now by orienting the coordinates  in the direction of the gap-averaged flow, say locally along $\mathbf{e}_s$, there will be only a single non-zero component of the velocity and pressure gradient. The velocity profile is easily found:
\begin{equation}\label{eq:velocity}
\hat{u}_s = \left\{
\begin{array}{ll}
- \displaystyle \frac{n}{1+n} \frac{1}{|\hat{\nabla}\hat{p}|^2 \hat{\kappa}^{\frac{1}{n}}} \left[  \left( |\hat{\nabla}\hat{p}| \hat{H} - \hat{\tau}_y \right)^{1+\frac{1}{n}} - \left( |\hat{\nabla}\hat{p}| \vert \hat{z} \vert - \hat{\tau}_y \right)^{1+\frac{1}{n}} \right] \frac{\partial \hat{p}}{\partial \hat{s}} & \mbox{iff}\quad \vert \hat{z} \vert > \displaystyle \frac{\hat{\tau}_y}{|\hat{\nabla}\hat{p}|}, \\[2pt]
\displaystyle  - \frac{n}{1+n} \frac{1}{|\hat{\nabla}\hat{p}|^2 \hat{\kappa}^{\frac{1}{n}}} \left( |\hat{\nabla}\hat{p}|\hat{H} - \hat{\tau}_y  \right)^{1+\frac{1}{n}} \frac{\partial \hat{p}}{\partial \hat{s}} & \mbox{iff}\quad \vert \hat{z} \vert \leqslant \displaystyle \frac{\hat{\tau}_y}{|\hat{\nabla}\hat{p}|} ,
\end{array} \right.
\end{equation}
which is the plane Poiseuille flow for a Herschel-Bulkley fluid. On integrating across the half-gap:
\begin{equation}
\label{eq:arealflow}
 |\hat{\nabla}\hat{\psi}| =
\hat{H} \hat{\bar{u}}_s = \int_0^{\hat{H}}\hat{u}_s ~\text{d} \hat{z} =
\frac{\displaystyle{ n \left( |\hat{\nabla}\hat{p}| \hat{H} - \hat{\tau}_y \right)^{1+\frac{1}{n}}_+ \left( (n+1) |\hat{\nabla}\hat{p}| \hat{H} + n \hat{\tau}_y  \right) } }{(n+1)(2n+1)\hat{\kappa}^{1/n} |\hat{\nabla}\hat{p}|^2}  .
\end{equation}
Here $(\cdot )_+$ denotes the positive part of the bracketed expression and note that $|\hat{\nabla}\hat{p}| = -\frac{\partial \hat{p}}{\partial \hat{s}}$. The combination $|\hat{\nabla}\hat{p}| \hat{H}$ that appears above is simply the wall shear stress $\hat{\tau}_w$. We observe that the flow stops (locally), provided that  $|\hat{\nabla}\hat{p}| \hat{H} \le \hat{\tau}_y $, showing that the flow is of limiting pressure gradient type.

While (\ref{eq:arealflow}) gives the algebraic relationship between the areal flow rate and the pressure gradient, we also want to use the inverse of this relation, which we write as
\[ |\hat{\nabla}\hat{p}| = \hat{S}( |\hat{\nabla}\hat{\psi}| ). \]
The function $\hat{S}( |\hat{\nabla}\hat{\psi}| )$ can be found numerically, provided that $|\hat{\nabla}\hat{\psi}| >0$, i.e.~by inverting the explicit function (\ref{eq:arealflow}).  The value of $\hat{S}$ as $|\hat{\nabla}\hat{\psi}| \to 0$, is the limiting pressure gradient: $\hat{\tau}_y /\hat{H}$. Subtracting off this limiting value, we can write:
\begin{equation}\label{eq:Sdefine}
  \hat{S}( |\hat{\nabla}\hat{\psi}| ) = \frac{\hat{\tau}_y}{\hat{H}} + \hat{\chi}(|\hat{\nabla}\hat{\psi}| ) =  \frac{\hat{\tau}_y}{\hat{H}} +  \frac{\hat{\tau}_w(|\hat{\nabla}\hat{\psi}| ) -  \hat{\tau}_y}{\hat{H}}.
\end{equation}
The first expression follows \cite{Bittleston2002}, whereas the second (equivalent) expression favours a description in terms of the wall shear stress \cite{Maleki2017}. The properties of $\chi(|\hat{\nabla}\hat{\psi}|$ are analysed in \cite{Pelipenko2004a}, wherein also various results on existence and uniqueness of solutions are developed.

Once  $\hat{S}( |\hat{\nabla}\hat{\psi}| )$ has been computed, we may eliminate the pressure as follows:
\begin{equation}\label{eq:streamfunctioneqn0}
  0 = \hat{\mathbf{\nabla}} \cdot \left( \frac{\partial \hat{p}}{\partial \hat{y}} , -\frac{\partial \hat{p}}{\partial \hat{x}} \right) = \hat{\mathbf{\nabla}} \cdot \left( \hat{S}( |\hat{\nabla}\hat{\psi}| ) \frac{\hat{\nabla} \hat{\psi} }{|\hat{\nabla} \hat{\psi}|}\right) .
\end{equation}

\subsection{Scaled problem}

For the scaled problem, we first scale the coordinates:
\[ \left( x,y \right) = \frac{\left( \hat{x},\hat{y} \right)}{\hat{L}}~~\mbox{ and }~~z=\frac{\hat{z}}{\hat{H}_0}  . \]
We suppose that the far-field pressure is $0$ and at the inflow $\hat{p} = \hat{p}_0$ is imposed, used to scale the pressure. The velocity scale $\hat{V}_0$ is used for velocities in the $\left( \hat{x},\hat{y} \right)$ plane and $\hat{H}_0 \hat{V}_0$ for the stream function. A timescale for the filling: $\hat{t}_0 = \hat{L}/\hat{V}_0$, is used for time, which only appears in the concentration equation.

Dimensionless variables use the same symbols as their dimensional equivalents, but without the $\hat{\cdot}$ symbol. Equation (\ref{eq:arealflow}) becomes:
\begin{equation}
\label{eq:arealflow_nd}
 |\nabla \psi| =
\frac{\displaystyle{ n \left( H |\nabla p| - Y \right)^{1+\frac{1}{n}}_+ \left( (n+1)H |\nabla p| + n Y \right) } }{(n+1)(2n+1) \kappa^2 |\nabla p|^2}  ,
\end{equation}
where
\begin{equation}
\kappa = c + (1-c) \mu_p:~~~~~~ \mu_p =  \frac{\hat{\kappa}_p }{\hat{\kappa}_s} \frac{\hat{V}_0^{n_p-n_s} }{\hat{H}_0^{n_p-n_s}} \ll 1  .
\end{equation}
The wall shear stress is $H |\nabla p|$ and $Y = \hat{\tau}_y  \hat{L} /(\hat{p}_0\hat{H}_0 )$. Note the limiting pressure gradient is now $Y/H$ and the scaled pressure gradient term is $|\nabla p| = S(|\nabla \psi|)$. The dimensionless viscous part of the pressure gradient is $\chi(|\nabla \psi|) = S(|\nabla \psi|) - Y/H$, which is obtained from (\ref{eq:arealflow_nd}):
\begin{equation}
\label{eq:arealflow_nd2}
 |\nabla \psi| = \frac{\displaystyle{ n \left( H \chi \right)^{1+\frac{1}{n}} \left( (n+1) H\chi + (n+2) Y \right) } }{(n+1)(2n+1) (\chi + Y/H)^2}  .
\end{equation}

After scaling, the stream function satisfies:
\begin{equation}\label{eq:streamfunctioneqn1}
  0 = \mathbf{\nabla} \cdot  \mathbf{S}: ~~~~~~  \mathbf{S} = \left[ \chi(|\nabla \psi|) + \frac{Y}{H} \right] \frac{\nabla\psi}{|\nabla\psi|} .
\end{equation}
The time advance is via the fluid concentration, which satisfies:
\begin{equation}\label{eq:conceqn1}
\frac{\partial}{\partial t}[ H \bar{c} ] +  \mathbf{\nabla} \cdot (H \bar{c}\bar{u}, H \bar{c}\bar{v} ) = 0 .
\end{equation}

\section{Model parameters and scope of study}

For our dimensionless model we essentially explore invasion (displacement) flows driven by unit pressure drop, where the in situ fluid is a low viscosity Newtonian fluid. The main resistance comes from the cement slurry and the limiting pressure gradient activates when the wall shear stress is less than the yield stress. This balance is captured in the scaled yield stress parameter $Y$. We expect that for large enough $Y$ the penetrating fluid will be unable to penetrate to the boundaries of the domain. The other parameters $\kappa$ and $n$, mostly influence the speed of the flow via the effective viscosity, i.e.~how fast is the steady flow achieved, but cannot stop the flow. The other key parameter is the microannulus geometry.

Our study is mostly focused on the establishment of a viable method for assessing the effectiveness of squeeze cementing operations, which depends on the degree of blockage of the microannulus after the cement slurry has been pumped. To do this we select the microannulus thickness using a stochastic model based on \cite{Trudel2023}; see \S \ref{sec:MAM}. We use the geometry to study two types of invasion problem in this paper: a planar invasion and a radial (perforation squeeze) invasion. Since generally the slurry is more viscous than the in situ fluid, we do not expect fingering type instabilities. With a uniform gap width, both problems would likely have a stable uniform  invasion. However, the irregular microannulus geometry ensures that the penetration is far from uniform.

\subsection{Planar invasion}

\begin{figure}[!h]
\centering
	\includegraphics[width=0.95\linewidth]{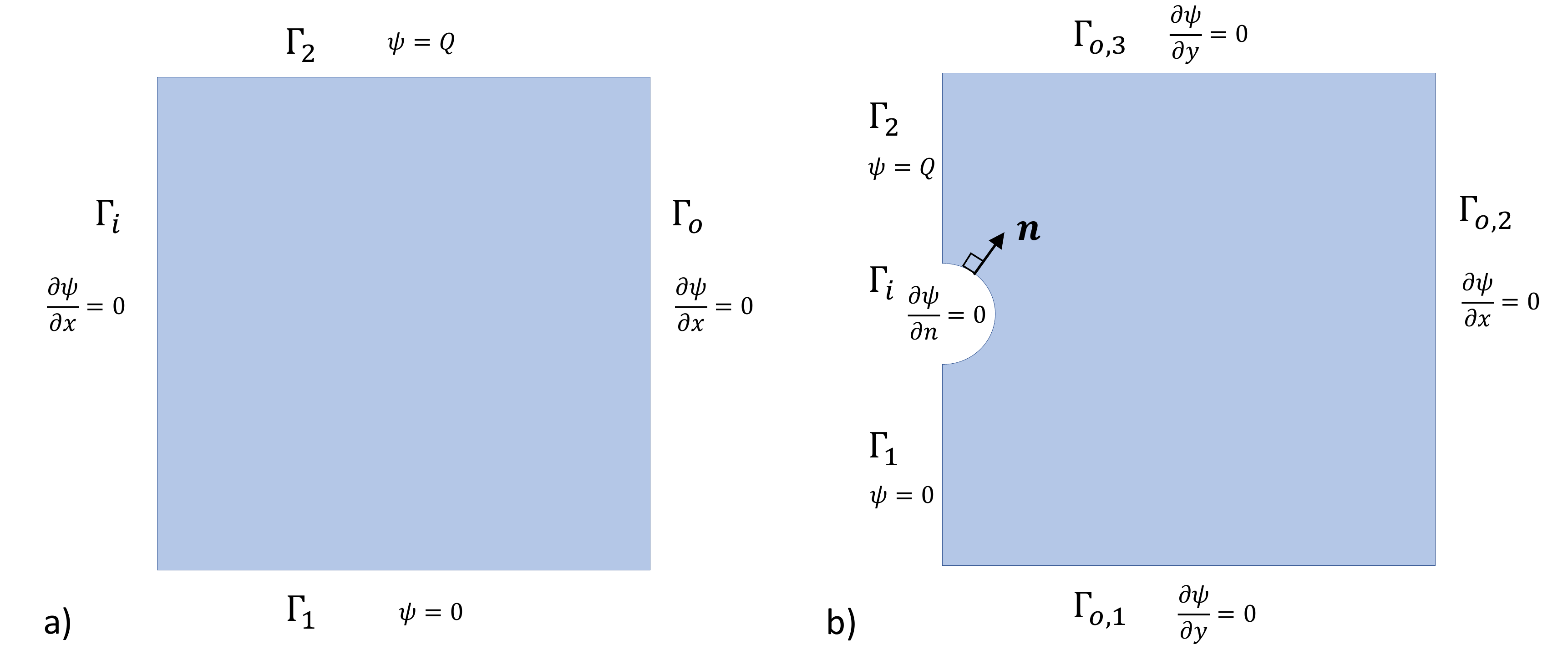}	
	\caption{Schematic of domains and boundary conditions for: a) planar invasion problem; b) perforation squeeze problem.}
	\label{fig:setup}
\end{figure}

Here we have a micro-annulus domain $\Omega = (x,y) \in (0,1) \times (-1/2,1/2)$, within which (\ref{eq:streamfunctioneqn1}) \& (\ref{eq:conceqn1}) are satisfied. The inflow/invasion occurs along $\Gamma_i$, at $x=0$, and the outflow/far-field is $\Gamma_o$, along $x=1$. The boundaries at $y = \mp 1/2$ are denoted $\Gamma_{1}$ \& $\Gamma_{2}$. No flow is allowed through $\Gamma_{1}$ \& $\Gamma_{2}$. Boundary conditions imposed are as follows; see Fig.~\ref{fig:setup}a).
\begin{itemize}
  \item Along $\Gamma_o$, we set the far-field pressure to $p=0$, which requires that
    \begin{equation}\label{eq:Gamma_o}
        \frac{\partial \psi}{\partial x}(1,y) = 0, ~~~~y \in  [-1/2,1/2] .
    \end{equation}
    Recall that the gradients of pressure and stream function are orthogonal and (\ref{eq:Gamma_o}) ensures constant $p$ along $\Gamma_o$.
  \item Along $\Gamma_i$, we would like to set the pressure $p=1$, meaning again that:
       \begin{equation}\label{eq:Gamma_i}
        \frac{\partial \psi}{\partial x}(0,y) = 0, ~~~~y \in  [-1/2,1/2] ,
    \end{equation}
to ensure a constant pressure along $\Gamma_i$.
\item Although  (\ref{eq:Gamma_o})  and  (\ref{eq:Gamma_i}) ensure a constant pressure on each boundary and consequently a constant pressure drop, there is no means of assuring that $p(0,y) - p(1,y) = 1$. This achieved via the upper and lower wall boundary conditions.  Along $\Gamma_{1}$ \& $\Gamma_{2}$, we impose:
    \begin{equation}\label{eq:Gamma_p}
	    \psi (x,-1/2) = 0,~~~~~~ \psi (x,1/2) = Q ~~~~x \in  [0,1].
    \end{equation}
    These conditions ensure a fixed flow rate $Q$ between the upper and lower boundaries, and that there is no flow through the boundaries.
\item  We iterate to find $Q$, such that
    \begin{equation}\label{eq:Gamma_i_p_}
	\int_{\Gamma_1} \frac{\partial p}{\partial x}dx = 1, ~~~\mbox{or}~~~\int_{\Gamma_2} \frac{\partial p}{\partial x}dx = 1 .
	\end{equation}
	Note that the flow rate monotonically increases with the pressure drop, enabling the solution to be found.
\item In advancing the concentration, the micro-annulus initially has $\bar{c}=0$ everywhere and $\tilde{c}=1$ is imposed along $\Gamma_i$. An outflow condition is used other boundaries when needed.
\end{itemize}

\subsubsection{Perforation squeeze}

For a perforation squeeze, we solve the equations (\ref{eq:streamfunctioneqn1}) \& (\ref{eq:conceqn1}) in a rectangular domain with a perforation hole in the centre. More generally this could be over a stencil of perforations, according to whatever shot pattern is implemented in the perforation gun. Here we are concerned mostly with feasibility. Thus, we consider a single perforation only and simplify to a half-domain, $\Omega = (x,y) \in (0,1/2) \times (-1/2,1/2)$, with semicircular inflow centred at $(0,0)$, with radius $r = 1/30$. We impose a symmetry condition along $x=0$. The inflow/invasion occurs radially outwards from $\Gamma_i$, where $\sqrt{x^2+y^2}=r$, and the outflow/far-field boundaries are $\Gamma_{o,1}$, $\Gamma_{o,2}$, \& $\Gamma_{o,3}$, defined as $y=1/2$, $x=1/2$, and $y=-1/2$, respectively; see Fig.~(\ref{fig:setup}b).

The micro-annulus domain initially has $c=0$ and for $t>0$, we impose $c=1$ along $\Gamma_i$ and outflow conditions elsewhere. Boundary conditions for (\ref{eq:streamfunctioneqn1}) are as follows.
\begin{itemize}
	\item Along $\Gamma_{o,1}$, $\Gamma_{o,2}$ and $\Gamma_{o,3}$, we have the far-field pressure $p=0$, which requires that
	\begin{equation}\label{eq:Gamma_o_radial}
	\begin{gathered}	
	\frac{\partial \psi}{\partial x}(1,y) = 0, ~~~~y \in  [-1/2,1/2],\\
		\frac{\partial \psi}{\partial y} (x,\pm 1/2) = 0, ~~~~x \in  [0,1/2].
	\end{gathered}
	\end{equation}
	The three equations in \ref{eq:Gamma_o_radial} imply a constant pressure along $\Gamma_{o,1}$, $\Gamma_{o,2}$ and $\Gamma_{o,3}$, and as they are connected, the pressure is equal on all these boundaries.
	\item Along $\Gamma_i$, we would like to set the pressure $p=1$. As in the planar flow we implement a constant pressure on this boundary, via the Neumann condition:
	\begin{equation}
		\frac{\partial \psi}{\partial n} = 0   ~~~~~\sqrt{x^2+y^2}=r,
	\end{equation}
	where $\mathbf{n}$ is the unit normal to $\Gamma_i$.
	\item As previously, we need to control the pressure drop between $\Gamma_i$ and the outflow. We do this via imposing a flow rate condition on $\Gamma_{1}$ and $\Gamma_{2}$:
	\begin{equation}
		\psi(0,y) = 0, ~~~~y \in [-1/2,-r],~~~~~~~~\psi(0,y) = Q, ~~~~y \in [r,1/2] .
	\end{equation}
    Here $Q$ represents the volumetric flowrate into the microannulus domain. Note that since $\psi$ is constant along $\Gamma_{1}$ and $\Gamma_{2}$, there is no flow across these boundaries (in agreement with the imposed symmetry). We now vary $Q$ in order to iteratively satisfy the pressure condition $p=1$, i.e.~we satisfy,
	    \begin{equation}\label{eq:iteration_Q_radial}
		\int_{r}^{1/2} \frac{\partial p(x,0)}{\partial x}~dx = 1.
	\end{equation}
\end{itemize}

\subsection{Micro-annulus model}
\label{sec:MAM}

Finally, we summarize the construction of a representative microannulus thickness using a stochastic model. As discussed in \S \ref{sec:intro}, characterization of microannulus geometry has been studied by a range of authors. Most studies point to highly varied microannulus thickness in the azimuthal direction while a certain continuity is noted in the well axis direction. The magnitude of these variation is not well known. The microannulus thickness typically range from 0 to 200 $\mu m$. While microannulus thickness of up to 1000 $\mu m$ have been measured, these are relatively uncommon. The distribution of microannulus thickness one might encounter is best described using right-skewed distributions as detailed by \cite{GarciaFernandez2019}. Based on these observations, in \cite{Trudel2023} we developed a stochastic model for microannulus thickness variation and calibrated the model against BC wellbore leakage data. We describe this model below.

\begin{figure}[!h]
\centering
		\includegraphics[width = 0.75\textwidth]{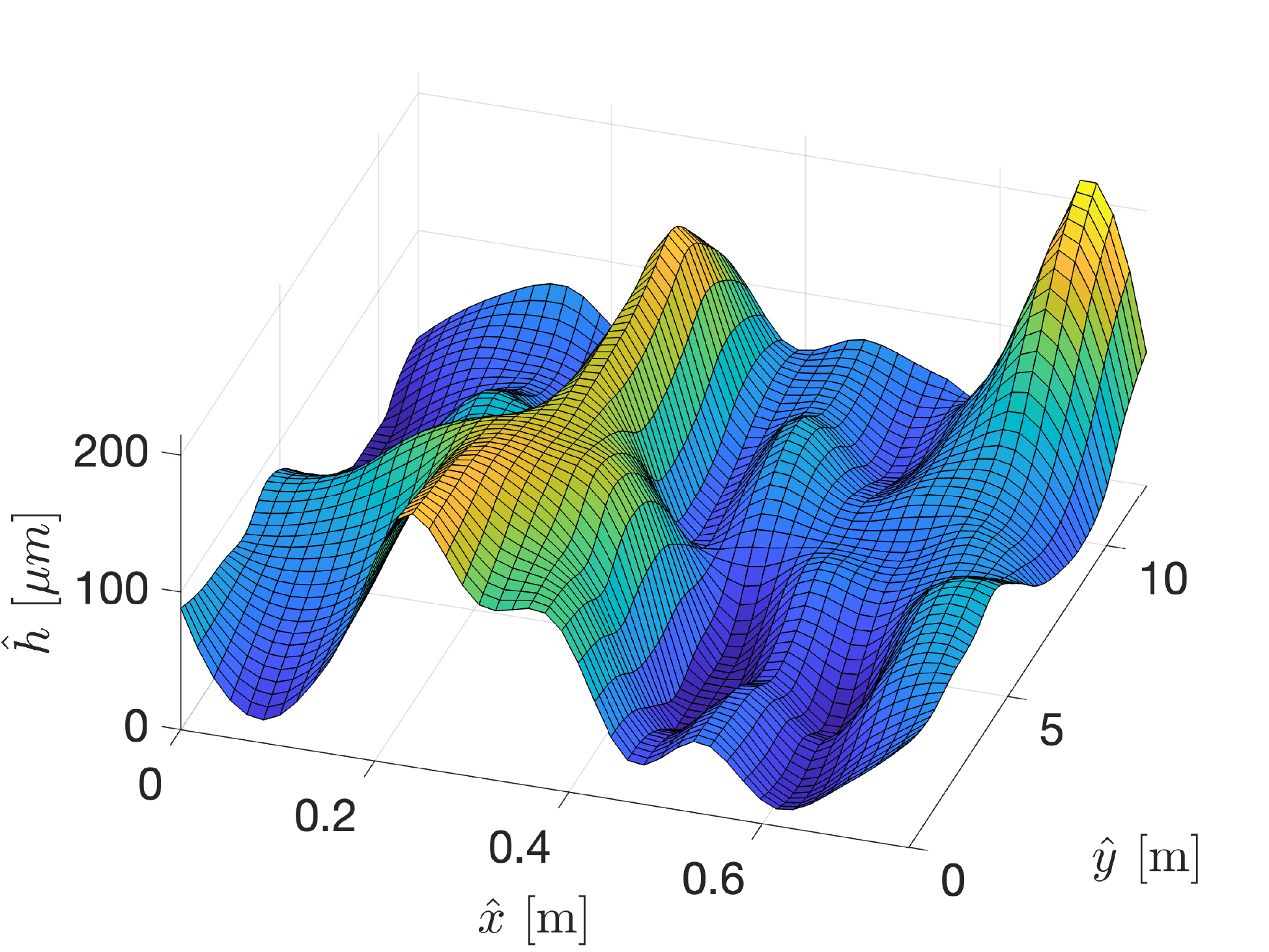}
		\caption{Microannulus geometry that results in a leakage rate of $2.00 m^{3}/day$, using the leakage model in \cite{Trudel2023}.}
		\label{fig:microannulus_geom}
\end{figure}

We construct $\hat{h}(\hat{x},\hat{y})$ over a rectangular domain. A nominal length $\hat{L}_c$, is used to represent axial variations along the well, in the direction $\hat{y}$, which is set to the length of a stand of casing used in industry. The coordinate $\hat{y}$ is oriented in the azimuthal direction. Values of $\hat{h}(\hat{x},\hat{y})$ are assigned on a mesh $(\hat{x}_i,\hat{y}_j)$ and used via interpolation. The microannulus thickness at each end of this section, at $\hat{y}= 0,\,\hat{L}_c$, are set based on random sampling of a lognormal distribution with parameters $\mu = 4.1,~\sigma = 1.15$ (with microannulus thickness measured in $\mu m$). These end values are then used to populate the microannulus thickness at interior points, with $\hat{h}(\hat{x},\hat{y})$ made periodic in $\hat{x}$. The interior points satisfy the following anisotropic averaging (diffusion) law:
\begin{equation}
\hat{h}_{i,j} = \frac{1}{2}m_{y}(\hat{h}_{i,j+1}+\hat{h}_{i,j-1}) +  \frac{1}{2}m_{x}(\hat{h}_{i+1,j}+\hat{h}_{i-1,j}) + \frac{\hat{\varepsilon}_{i,j}}{\sqrt{2}} ,
 \label{eq:microannulus_geom}
\end{equation}
where $\hat{\varepsilon}_{i,j}$ is a random variable sampled from a normal distribution with zero mean and standard deviation $a \delta \hat{y}$. Here $a$ controls the dimensionless amplitude of the stochastic perturbation to the averaging operator and $\delta \hat{y}$ the mesh size used in $\hat{y}$. Provided the mesh sizes in $(\hat{x},\hat{y})$ directions have a fixed ratio, including $\delta \hat{y}$ in the amplitude of $\hat{\varepsilon}_{i,j}$ controls the mesh dependency.
The parameters $m_{x}$ and $m_y$ satisfy $m_x + m_y = 1$. The observed anisotropy in $\hat{h}(\hat{x},\hat{y})$, i.e. stronger variations azimuthally than axially, is controlled by selecting $m_{x} \ll m_y$. In \cite{Trudel2023} each microannulus geometry is used in a leakage model based on the typical parameters of a BC well, resulting in computation of a leakage rate; an example is given in Fig.~\ref{fig:microannulus_geom}.

\begin{figure}[!h]
	\centering
	\includegraphics[width=0.8\linewidth]{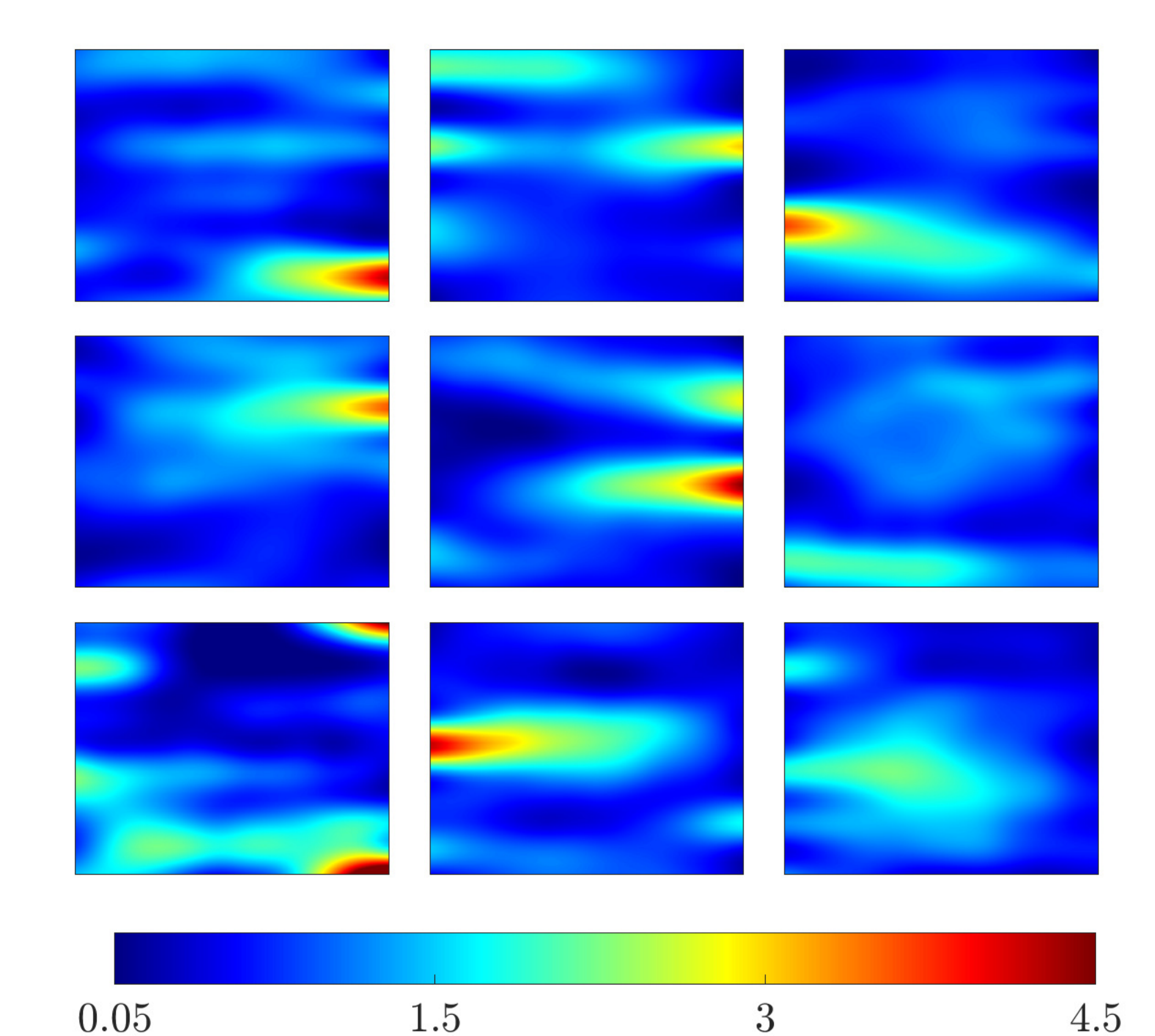}	
	\caption{Examples of stochastically generated gap widths $H(x,y)$, normalized to have an average equal to one.}
	\label{fig:heightSamples.eps}
\end{figure}

Here we are interested in the squeeze cementing flows that penetrate microannular spaces, rather than gas flow along a microannulus. Although we use the above model for microannulus thickness, the length-scale $\hat{L}_c$ is too long. In squeeze cementing the perforation holes are made in a pattern that involves spacing azimuthally and along the well, typically spaced at distances $\sim 10-40cm$. Thus, for our purposes we calculate  $\hat{h}(\hat{x},\hat{y})$ as above, sample a rectangle of appropriate size for the geometry we consider, average $\hat{h}$ to compute $\hat{\bar{h}} = 2\hat{H}_0$ which is used to normalize and define $H(x,y)$. A selection of normalised $H(x,y)$ are illustrated in Fig.~\ref{fig:heightSamples.eps}, which show the directional anisotropy and the typical variations.

\section{Numerical method }
\label{sec:numerical}

The numerical challenge is only to compute the velocity field (stream function) from (\ref{eq:streamfunctioneqn1}), as computation of the concentration equation is more standard. Our method is based on the variational formulation of (\ref{eq:streamfunctioneqn1}) and equivalent minimization, as introduced by \cite{Pelipenko2004b}. Here we outline the method focusing on the planar invasion flow, with the perforation flow being similar.
We formally define 2 test sets for the solution:
\begin{eqnarray}
  \mathcal{V} &=& \left\{ \phi \in W^{1,1+n}(\Omega): ~\phi = 0 \mbox{ on } \Gamma_1;~~ \phi = Q  \mbox{ on }  \Gamma_2  \right\} \\
 \mathcal{V}_0 &=& \left\{ \tilde{\phi} \in W^{1,1+n}(\Omega): ~\phi = 0 \mbox{ on } \Gamma_1;~~ \phi = 0  \mbox{ on }  \Gamma_2   \right\}
\end{eqnarray}
The Sobolev space $W^{1,1+n}(\Omega)$ is the general solution space; see \cite{Pelipenko2004b}. The space $\mathcal{V}$ includes the essential boundary conditions. For the variational setting and minimization we need a closed space; hence $\mathcal{V}_0$. Evidently $\mathcal{V}$ is not empty as for example we find $\psi_0 = Q(y+1/2) \in \mathcal{V}$,. For a given $\psi_0$, for any $\phi \in \mathcal{V}$ we have $\tilde{\phi} = \phi - \psi_0 \in \mathcal{V}_0$. Therefore, the optimization can be formally carried out over $\mathcal{V}_0$ and combined with $\psi_0$ for the solution in $\mathcal{V}$.

We denote the solution by $\tilde{\psi} \in \mathcal{V}_0$ and $\tilde{\phi}$ is any other test function in $\mathcal{V}_0$. From (\ref{eq:streamfunctioneqn1}), on using the divergence theorem:
\begin{equation}\label{eq:var0}
  0 = \int_{\Gamma_i \bigcup \Gamma_{1} \bigcup \Gamma_{o} \bigcup \Gamma_{2} }(\tilde{\phi}-\tilde{\psi})  \mathbf{S} \cdot \mathbf{n}~d\Gamma  - \int_{\Omega} \mathbf{S} \cdot \nabla(\tilde{\phi}-\tilde{\psi}) ~d\Omega.
\end{equation}
The conditions on $\mathcal{V}_0$ ensure that the integrals along $\Gamma_{1}$ \& $\Gamma_{2}$ vanish. The integrals along $\Gamma_{i}$ \& $\Gamma_{o}$ vanishes only provided $S_x = 0$ for the solution. This is satisfied since $p$ is constant and $S_x = \frac{\partial p}{\partial y} = 0$. With some further manipulations we find that (\ref{eq:var0}) becomes
\begin{equation}\label{eq:var1}
\int_{\Omega} \frac{\chi(|\nabla \psi|)}{|\nabla\psi|} \nabla\psi \cdot \nabla(\tilde{\phi}-\tilde{\psi}) + \frac{Y}{H} (|\nabla(\psi_0 + \tilde{\phi})| - |\nabla(\psi_0 + \tilde{\psi})| )~d\Omega  \ge 0, ~~\forall  \tilde{\phi} \in \mathcal{V}_0  .
\end{equation}
The elliptic variational inequality above is equivalent to the following minimization problem:
\begin{equation}\label{eq:var2}
\min_{\tilde{\phi} \in \mathcal{V}_0 } J(\nabla \tilde{\phi}),~~~~~J(\nabla \tilde{\phi}) =
\int_{\Omega} \left( \int_0^{|\nabla(\psi_0 + \tilde{\phi})|} \chi(a)~da + \frac{Y}{H} |\nabla(\psi_0 + \tilde{\phi})| \right)~d\Omega   .
\end{equation}
The minimization, equation (\ref{eq:var2}), is of a standardised form. The non-differentiability of the second term in $J$ is dealt with either by regularization or by an augmented Lagrangian method (which we adopt here). This amounts to solving the following saddle point problem \citep{Pelipenko2004a,Pelipenko2004b,Maleki2017}:
\begin{equation}\label{eq:ALp1}
\min_{\tilde{\phi} \in \mathcal{V}_0,~\tilde{\mathbf{q}} \in [L^{1+n}(\Omega)]^2 } \max_{\mathbf{\mu} \in [L^{1+n}(\Omega)]^2 } \left\{ J(\tilde{\mathbf{q}}) + \int_{\Omega} \mathbf{\mu} \cdot (\nabla \tilde{\phi} - \tilde{\mathbf{q}} )~d\Omega + \frac{r}{2}\int_{\Omega} |\nabla \tilde{\phi} - \tilde{\mathbf{q}} |^2 ~d\Omega  \right\} .
\end{equation}
Here for the solution we generally have $\tilde{\phi} \to \tilde{\psi}$, and $\tilde{\mathbf{q}} \to \nabla \tilde{\psi}$. The variable $\mathbf{\mu} $ is the Lagrange multiplier for the constraint $\nabla \tilde{\phi} = \tilde{\mathbf{q}}$, and is found to converge to $\mathbf{S}$. We discuss the solution iteration steps below.

Note first that while the above saddle point finds $\psi$ for a given $Q$, the pressure constraint is not satisfied. This constraint can be satisfied either in an outer iteration, or as part of the iteration for solving (\ref{eq:ALp1}).

\subsection{Uzawa iteration}

The saddle point problem is solved by sequentially minimizing for each variable. We assume at the $k$-iterate that we have $\tilde{\psi}^k$, $\tilde{\mathbf{q}}^k$ and $\mathbf{\mu}^k$. Then $\tilde{\psi}^{k+1}$ satisfies:
\begin{equation}\label{eq:ALpsi1}
  0 = \int_{\Omega} \left( -r \nabla^2 \tilde{\psi}^{k+1} + r \mathbf{\nabla} \cdot \tilde{\mathbf{q}}^k - \mathbf{\nabla} \cdot \mathbf{\mu}^k \right) \tilde{\phi}~ d\Omega ~~~~~~\forall  \tilde{\phi} \in \mathcal{V}_0 .
\end{equation}
This is a linear Poisson equation at each step. Note that the non-essential conditions on $\Gamma_i$ and $\Gamma_o$ can enforced by assuring that $\tilde{q}^k_x = 0$ and $\mu_x^k = 0$ on these boundaries.

Note that $Q$ does not explicitly appear in this equation, but defines $\psi_0$. Generally \ref{eq:ALpsi1} requires a matrix system to be solved, for each new value of $r \mathbf{\nabla} \cdot \tilde{\mathbf{q}}^k - \mathbf{\nabla} \cdot \mathbf{\mu}^k$, but the same linear system is solved for successive iterates, which can allow significant speed up. Having computed $\tilde{\psi}^{k+1}$ one can calculate the pressure drop due to  $\tilde{\psi}^{k+1}$ and adjust $Q$ directly to produce the desired pressure drop, since the pressure drop is proportional to $Q$.

To find $\tilde{\mathbf{q}}^{k+1} \in [L^{1+n}(\Omega)]^2$, we need only minimize locally, e.g.~on each discrete element. It has more physical meaning to add $\nabla \psi_0$ to both $\tilde{\mathbf{q}}^{k+1}$ and $\nabla \tilde{\psi}^{k+1}$. We then find $\mathbf{q}^{k+1} = \nabla \psi_0 + \tilde{\mathbf{q}}^{k+1}$ as the minimizer of:
\begin{equation}\label{eq:ALq}
\min_{\mathbf{q}} \left\{ \int_0^{q} \chi(a)~da + \frac{Y}{H} q + \frac{r}{2}q^2 - (\mathbf{\mu}^k + \nabla \psi^{k+1} ) \cdot \mathbf{q}   \right\}.
\end{equation}
Here we have written $q = |\mathbf{q}|$. The minimizer must be parallel to  $\mathbf{m} = (\mathbf{\mu}^k + \nabla \psi^{k+1} )$, so we may write: $\mathbf{q} = q \mathbf{m} /m $. Equation (\ref{eq:ALq}) becomes a minimization over single variable $q$:
\begin{equation}\label{eq:ALq1}
\min_{q} \left\{ \int_0^{q} \chi(a)~da  + \frac{r}{2}q^2 + \left(\frac{Y}{H} - m \right) q  \right\}.
\end{equation}
Here $m$ plays the role of the pressure gradient. We see that if $Y \ge Hm$ then the solution is $q=0$. Otherwise, supposing that $Y < Hm$ we differentiate (\ref{eq:ALq1}) to find $q$ from:
\begin{equation}\label{eq:ALq2}
\chi(q)  + r q = m -\frac{Y}{H} > 0,
\end{equation}
which has a single solution since $\chi(q)$ increases strictly. To find this solution, the computation can be accelerated by using the algebraic relation (\ref{eq:arealflow_nd2}), which is the implicit definition of $\chi$. In this way, we still find the root of a monotone algebraic equation but do not need to evaluate $\chi$ explicitly at each step. More clearly, instead of   (\ref{eq:ALq1}) we solve  (\ref{eq:ALq3}) for $\chi$:
\begin{equation}\label{eq:ALq3}
\chi  + r  \frac{\displaystyle{ n \left( H \chi \right)^{1+\frac{1}{n}} \left( (n+1) H\chi + (n+2) Y \right) } }{(n+1)(2n+1) (\chi + Y/H)^2}  = m -\frac{Y}{H} > 0.
\end{equation}
We then also have:
\begin{equation}
\label{eq:arealflow_nd2_q}
q = \frac{\displaystyle{ n \left( H \chi \right)^{1+\frac{1}{n}} \left( (n+1) H\chi + (n+2) Y \right) } }{(n+1)(2n+1) (\chi + Y/H)^2}  \mbox{ and } \mathbf{q}^{k+1} = \mathbf{q} = q \frac{\mathbf{m}}{m}.
\end{equation}
Observe that since $\mathbf{m} = (\mathbf{\mu}^k + \nabla \psi^{k+1} )$, the flow rate $Q$ is explicitly included via $\psi_0$. As the iteration proceeds, $\mathbf{\mu}^k \to \mathbf{S}$, and the condition $Y \ge Hm$, becomes the limiting pressure gradient condition (as also $\nabla \psi^{k+1} \to 0$ in these regions).

Lastly, for $\mathbf{\mu}^{k+1}$ we use a projection method:
\begin{equation}\label{eq:ALmu}
  \mathbf{\mu}^{k+1} = \mathbf{\mu}^{k} + \varrho [\nabla \psi^{k+1} - \mathbf{q}^{k+1} ] .
\end{equation}

\subsection{Discretization overview}

For the implementation we adopt a finite element formulation to discretize the equations. We use a piece-wise quadratic continuous element (P2) for the streamfunction, $\psi$, and a piece-wise linear discontinuous element, (P1-DC), for other fields, i.e, $\mathbf{q},~\mathbf{\mu},~c$, and $H$. The sequential iteration of the solution of equations (\ref{eq:ALpsi1}), (\ref{eq:ALq1}) \& (\ref{eq:ALmu}) proceeds until
$$ || \tilde{\psi}^{n+1} - \tilde{\psi}^n || < tol= 10^{-4} Q . $$
Note that including $Q$ preserves the accuracy as the flow stops. This usually takes a few hundred iterations to converge. This is repeated for every time advance.

We used the dual-P1-DC formulation \cite{ern2006discontinuous}, for the concentration equation. This takes the following form:
\begin{equation}
\int_\Omega \left( \frac{H(c^{n+1}-c^{n})}{\delta t} +\bar{\mathbf{u}} \cdot \mathbf{\nabla} c \right)\omega ~d\Omega+ \int_E \left(0.5(|\mathbf{n}.\bar{\mathbf{u}}|-\mathbf{n}.\bar{\mathbf{u}}) \right) [c] \omega ~ds = 0 ~~~~\forall \omega,
\end{equation}
where $E$ is the set of internal edges, and $[c]$ denotes the jump of $c$ across an edge.

\subsection{Benchmark flow problems}

\begin{figure}[!h]
	\centering
	\includegraphics[width=1.2\linewidth]{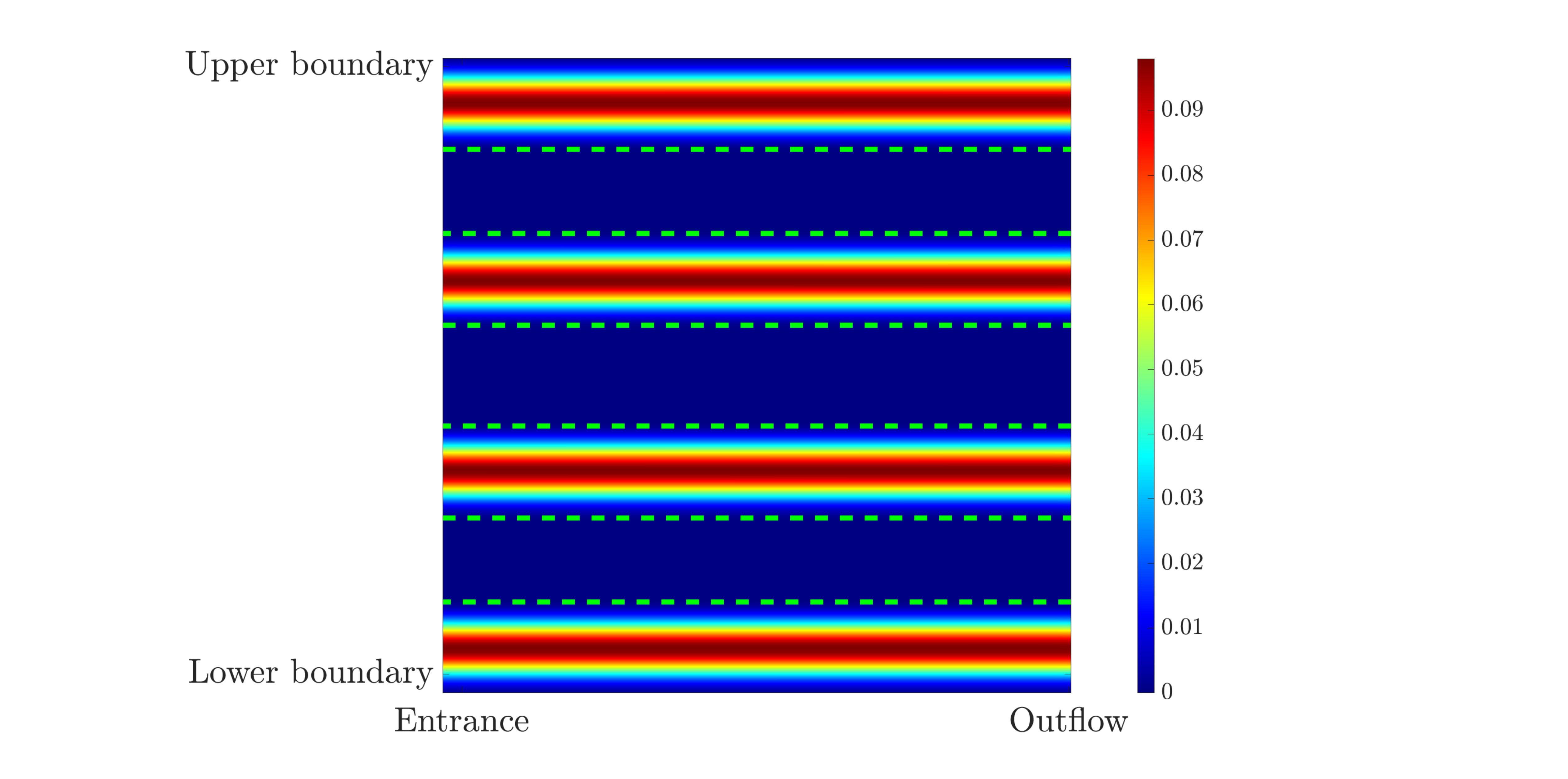}	
	\caption{The velocity magnitude for the flow of a viscoplastic fluid, $Y=0.5$, and $n=1$, under a constant pressure drop equal to one, between entrance (left hand side) and the outflow (right hand side) boundaries. The domain is similar to fig \ref{fig:setup}a ,i.e., $\Omega = (x,y) \in (0,1) \times (-0.5, 0.5)$}
	\label{fig:validation1:plug.eps}
\end{figure}

\begin{figure}[!h]
		\centering
	\includegraphics[width=0.8\linewidth]{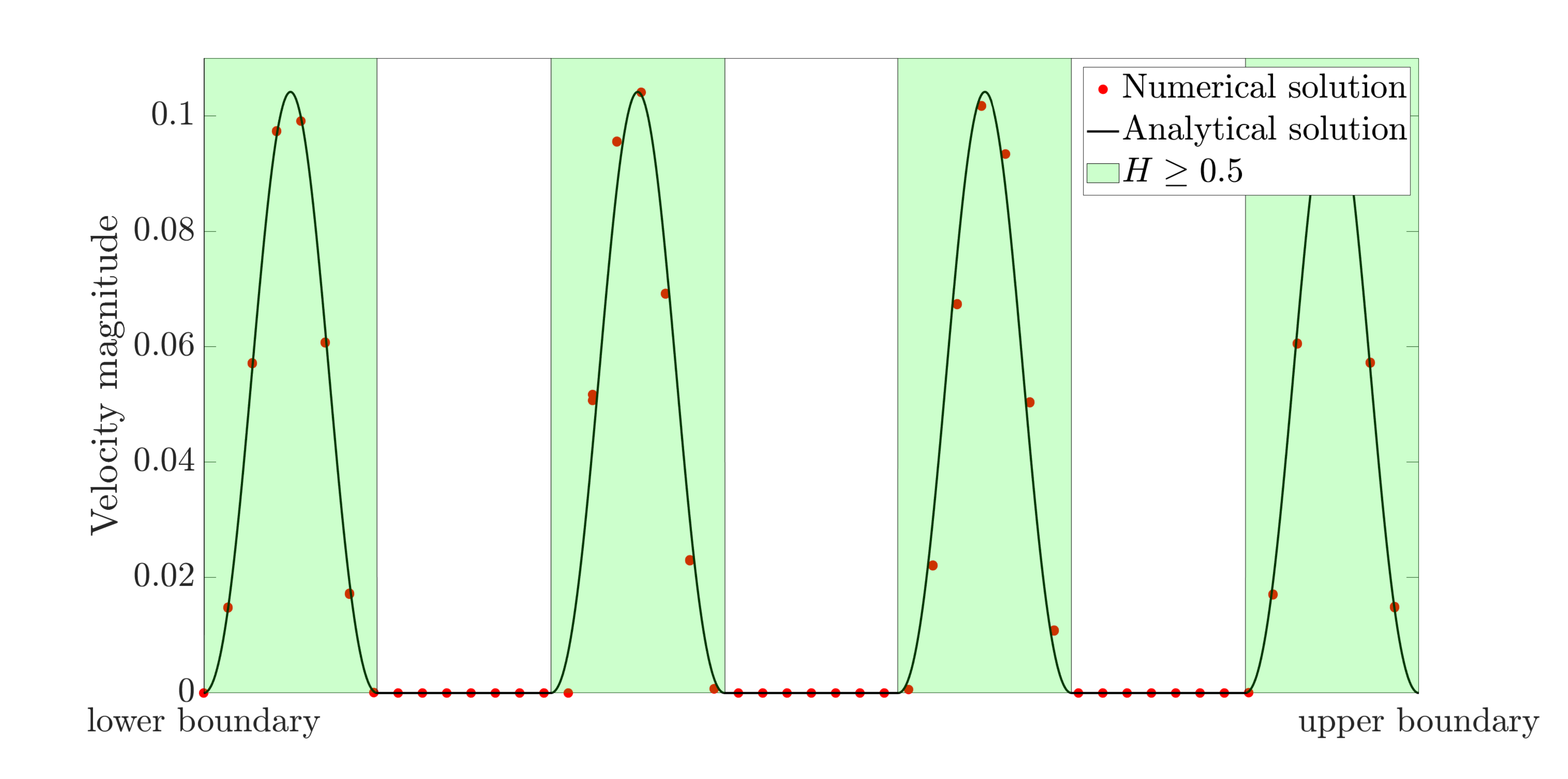}	
	\caption{Comparison of velocity magnitude from our numerical results (red dots) versus the analytical solution (black line). This is the velocity profile at the outflow shown in figure \ref{fig:validation1:plug.eps}.}
	\label{fig:validation1:velocity.eps}
\end{figure}

To benchmark, we investigate two single-phase flow problems. Firstly, we model the flow of a viscoplastic fluid, with $Y=0.5$ and $n=1$, along a Hele-Shaw cell with a domain $\Omega = (x,y) \in (0,1) \times (-0.5,0.5)$ under a constant pressure drop equal to one. The Hele-Shaw cell has a thickness $H(y)$, periodically varying in $y$. We adopt a sinusoidal variation as:
\begin{equation}
	H(y)=\frac{\sin(7 \pi y)+1}{2}.
\end{equation}
The fluid flows in one direction only, along the Hele-Shaw cell. The flow velocity at each $y$ can be determined via solution of a single nonlinear algebraic equation, i.e.~equation (\ref{eq:arealflow_nd2}). Figure (\ref{fig:validation1:plug.eps}) shows the velocity magnitude in this domain. We compared the results with the analytical solution, computed by evaluating (\ref{eq:arealflow_nd}) for a unit pressure gradient. Figure (\ref{fig:validation1:velocity.eps}) shows a comparison of the analytical solution (solid black line) and the numerical  results (red dots).

\begin{figure}[!h]
			\centering
\includegraphics[width = \linewidth]{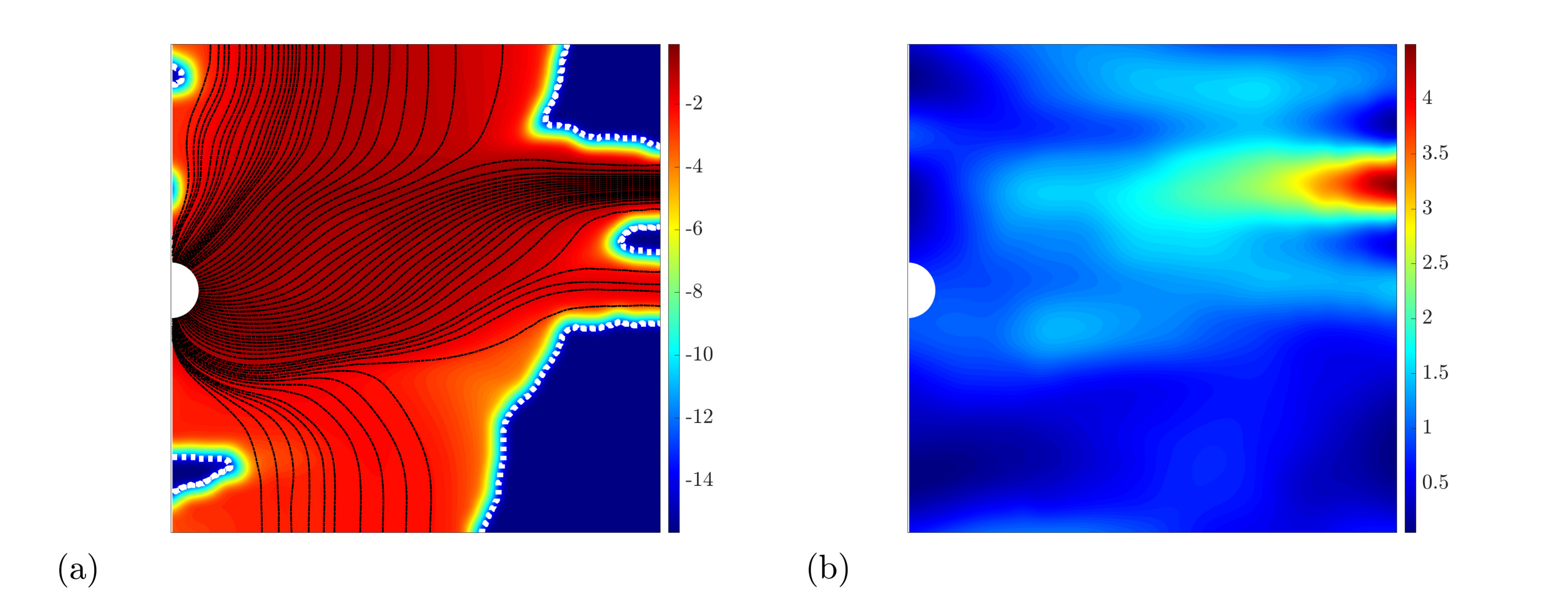}	
    \caption{Example single phase flow through perforation hole: (a) Logarithm of the velocity magnitude computed in the randomized geometry. The white dashed lines separates the yielded and un-yielded zones, and the black lines represent the streamlines. Parameters are $Y=0.5$ and $n=1$; (b) Randomized $H$ for $\Omega = (x,y) \in (0,0.5) \times (-0.5, 0.5)$.}
\label{fig:validation2:loqq_n.eps}
\end{figure}

The second problem considered is a single-phase flow through a randomized $H$. Figure \ref{fig:validation2:loqq_n.eps}(a) shows the variation of $H$ and Fig.~\ref{fig:validation2:loqq_n.eps}(b) shows the velocity magnitude in the domain, corresponding to a fixed unit pressure drop between inflow and outflow. The white dashed lines separate the unyielded areas from the flow domain. The black lines represent the streamlines. As can be seen, the flow finds a preferential path through areas with higher $H$. The domain is similar to Fig.~\ref{fig:setup}b, i.e.~$\Omega = (x,y) \in (0,0.5) \times (-0.5, 0.5)$, and a constant pressure drop equal to one is applied between the injection hole and the outer boundaries.

\section{Results: invasion problems}

We now turn to invasion problems. Our simulations are run until a final \emph{stoppage time}, at which the $\ell^2$-norm of the difference of concentration between two consecutive time steps falls below a set tolerance:
$$||c^{n+1}-c^n|| < 10^{-4} . $$
In reality the pumping operation continues until a the pump pressure rises sharply, when the flow is reduced in order to prevent any fracturing, (called a low pressure squeeze).

\subsection{Planar invasion}

We first look at a benchmark planar invasion problem where a viscoplastic fluid, $Y=0.5$ and $n=1$, displaces a Newtonian fluid in the same corrugated geometrical setup as the example Fig.~\ref{fig:validation1:plug.eps}. We expect the viscoplastic fluid to follow  preferential pathways where $H$ is largest. This is indeed the case and the main observation is that the large disparity in velocity leads to some dispersion of the intermediate values of $c$. Secondly, due to the relatively large $Y$ values, a part of the channel remains immobile, where $H$ is small, i.e.~the pressure gradient is $\approx -1$ and therefore, when $H \approx H|\nabla p| < Y$ there should be zero flow.

\begin{figure}[!h]
\centering
	\includegraphics[width=\linewidth]{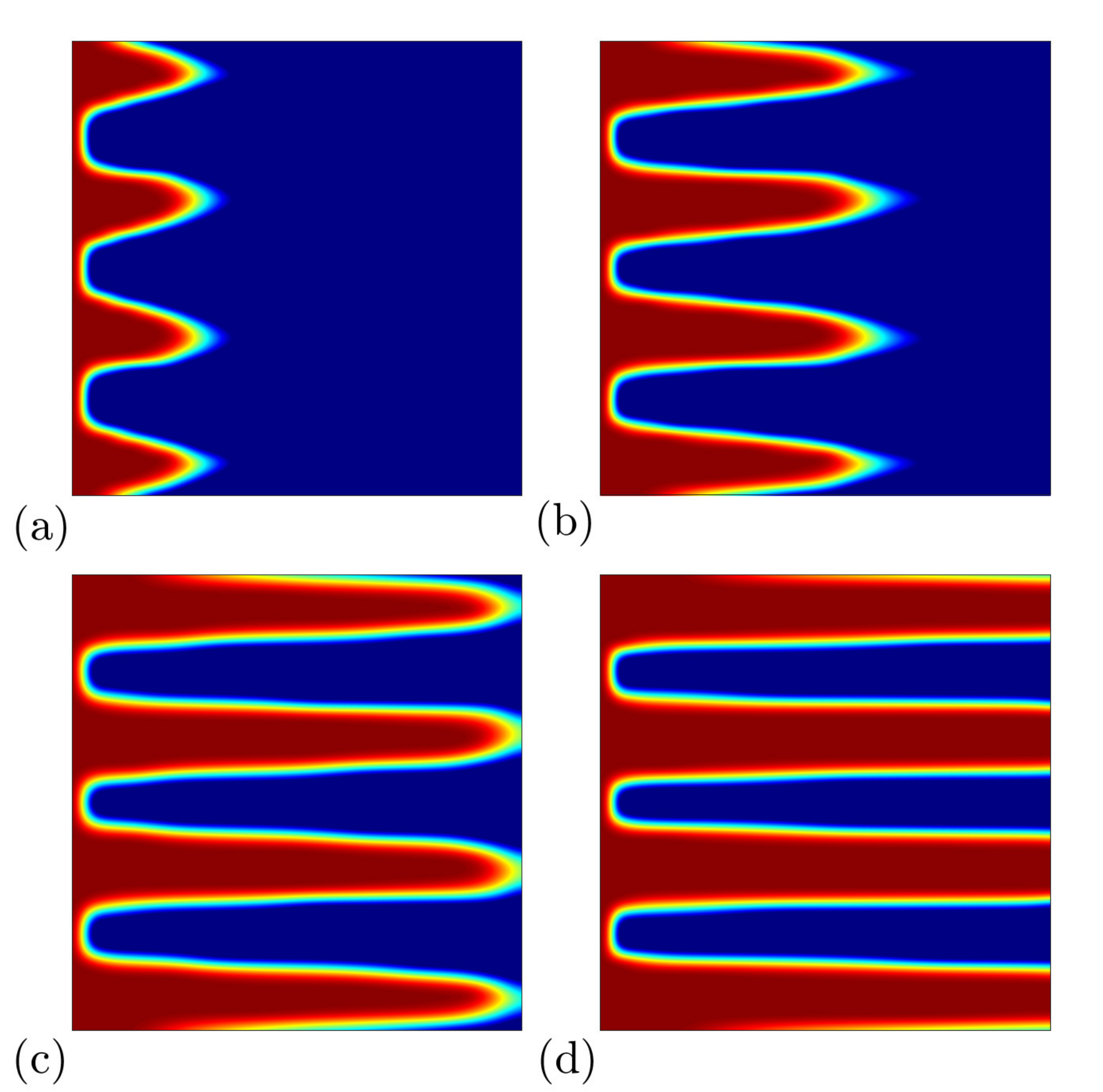}	
	\caption{The concentration of the displacing fluid, $Y=0.5,~n=1$, in four snapshots. Each snapshot shows the time equal to 1/4, 2/4, 3/4 and 4/4 of the convergence time ($\approx$ 213.8). The domain is similar to fig \ref{fig:setup}a ,i.e., $\Omega = (x,y) \in (0,1) \times (-0.5, 0.5)$, and a constant pressure drop equal to one is applied between entrance and outflow.}
	\label{fig:validation3:conc.eps}
\end{figure}

We next investigate planar invasion in a randomized geometry, shown at 4 successive timeslots during the run. By looking at the white streamlines, it can be seen that the preferential paths may change as the invasion advances. The yield stress of the displacing fluid, stops the flow locally wherever the pressure gradient is too small. In addition, as the flow progresses, because a larger portion of any pathway is filled with cement, the resistance against flow increases, making the unfilled pathways easier to flow along.

\begin{figure}[!h]
\centering
	\includegraphics[width=\linewidth]{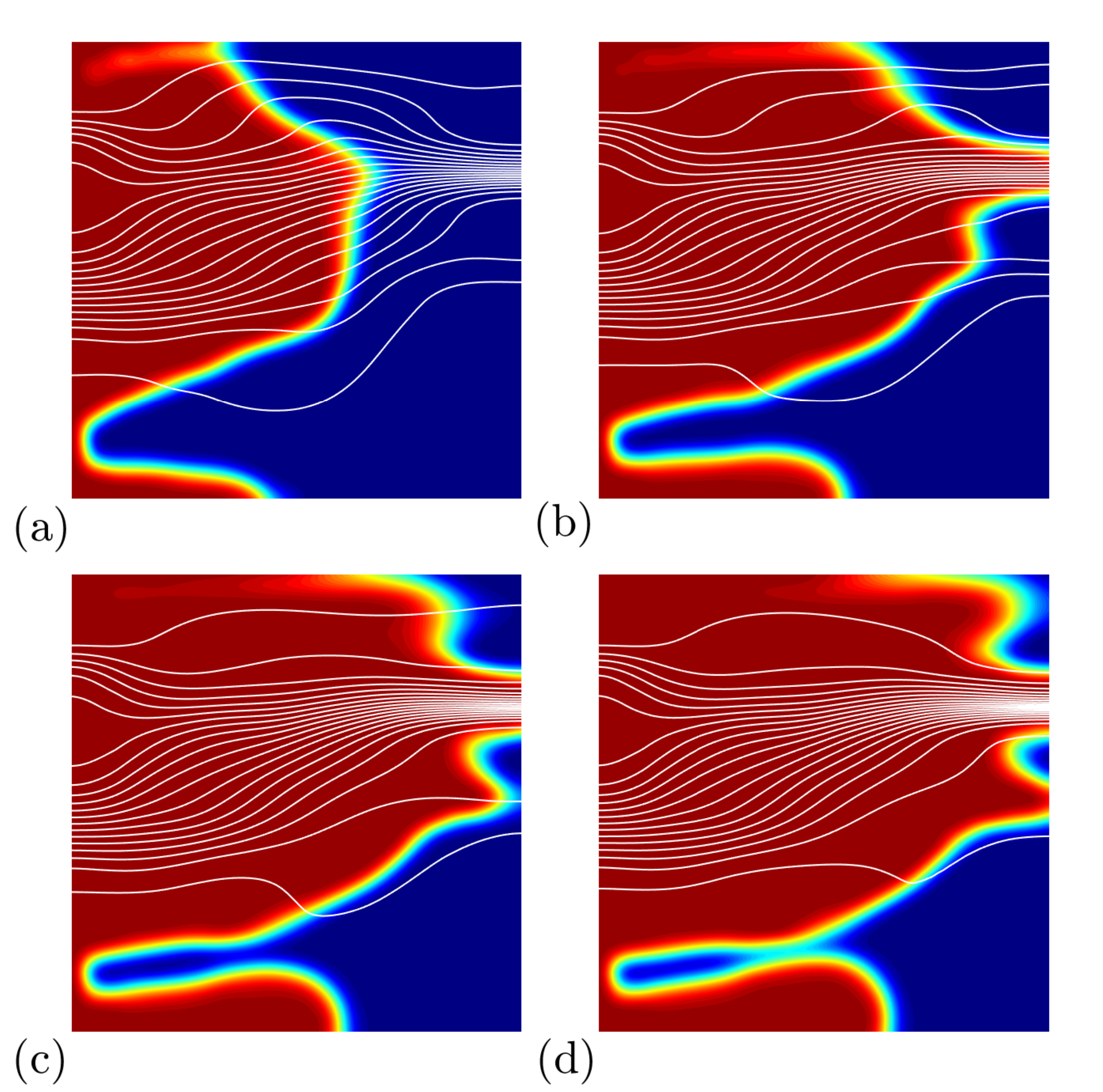}	
	\caption{The concentration of displacing fluid $Y=0.5,~n=1$, in four consecutive time steps, $t=1/4,~ 2/4,~ 3/4$, and $4/4$ of the convergence time ($\approx21.7$). The white lines show the streamlines. The domain is similar to fig \ref{fig:setup}A ,i.e., $\Omega = (x,y) \in (0,1) \times (-0.5, 0.5)$, and a constant pressure drop equal to one is applied between Entrance and outflow.}
	\label{fig:axial.eps}
\end{figure}

\subsection{Invasion from a perforation}

We now consider radial displacement with the setup of Fig.~\ref{fig:validation1:plug.eps}(b). The aim is to explore the potential of the computational simulation for calculating different metrics that may be of help in deciding whether the squeeze flow is successful or not. We start with 3 examples taken from 3 different random geometries. In each we compute the invasion flow, displacing the Newtonian fluid with a viscoplastic fluid: $n=1$ and 3 yield numbers, $Y=0.5,~1,~1.5$. In each figure, image A shows the microannulus thickness $H$. Images B-D show the simulation results for the 3 yield numbers. In each of these, the colourmap corresponds to the final stoppage time. The dashed lines mark $\bar{c}=0.5$ at the stoppage time as well as at $1/3$ \& $2/3$ of the stoppage time.

\begin{figure}[!h]
	\includegraphics[width=\linewidth]{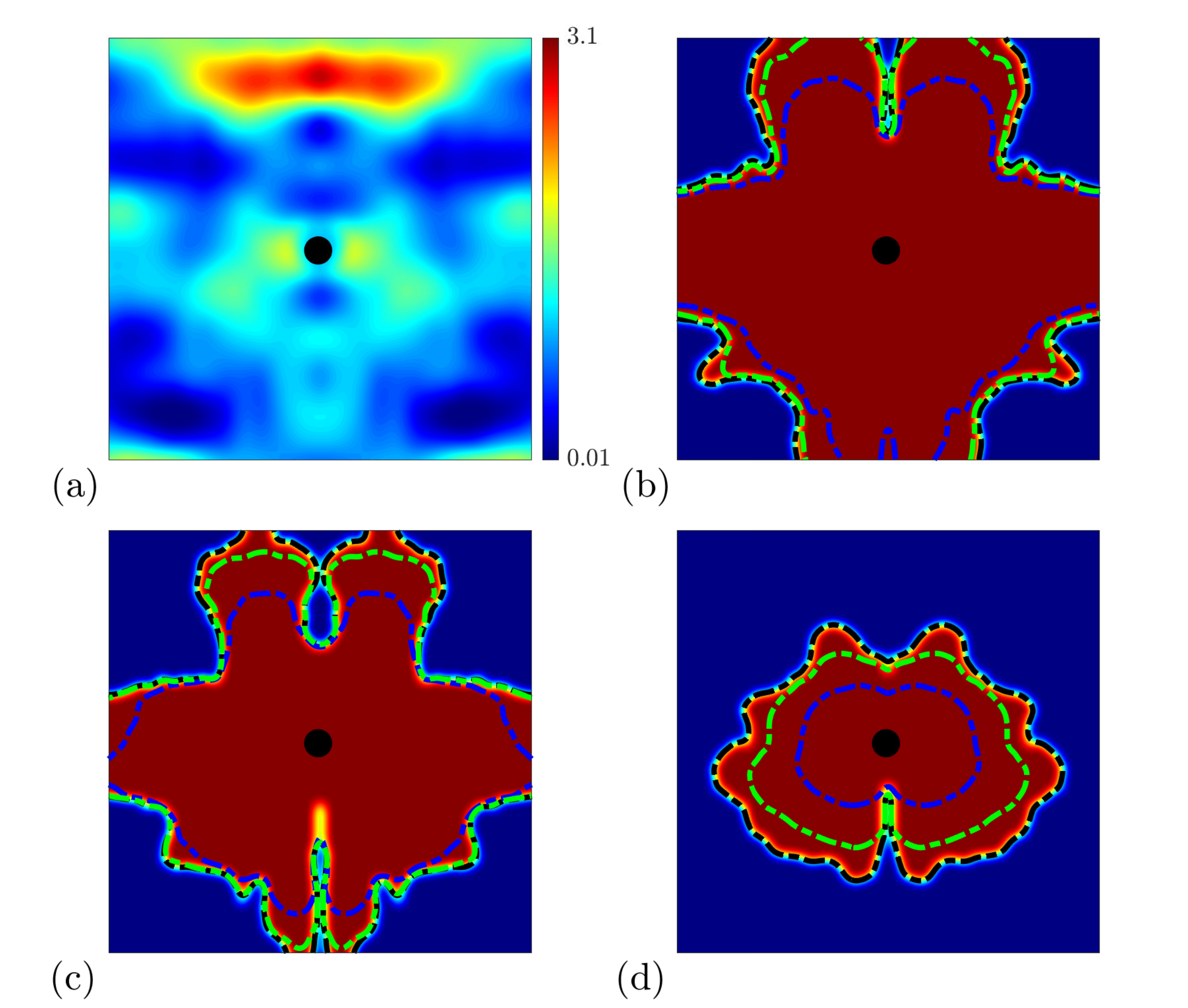}	
	\caption{The front evolution during displacement of a Newtonian fluid by a Bingham fluid: (b) $Y=0.5$; (c) $Y=1$; (d) $Y=1.5$; all flows through the randomized geometry shown in panel (a). The blue, green, and black dashed lines show the displacing front, where $\bar{c}=0.5$, at 1/3, 2/3 and 1 of the stoppage time.}
	\label{fig:radial:1st.eps}
\end{figure}

This example shows the attenuation and omission of one of the paths exiting from the lower boundary.

\begin{figure}[!h]
	\includegraphics[width=\linewidth]{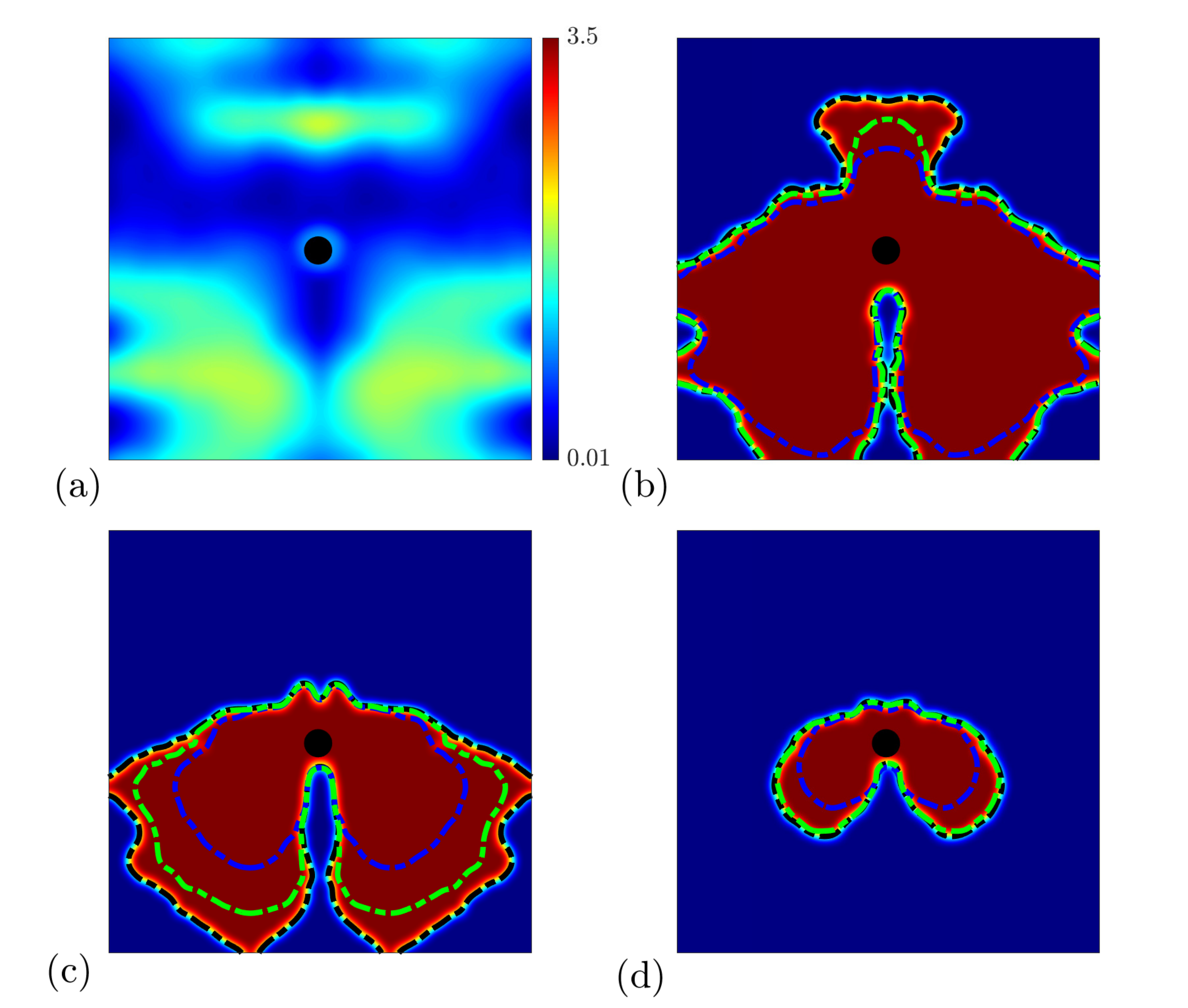}	
	\caption{The front evolution during displacement of a Newtonian fluid by a Bingham fluid: (b) $Y=0.5$; (c) $Y=1$; (d) $Y=1.5$; all flows through the randomized geometry shown in panel (a). The blue, green, and black dashed lines show the displacing front, where $\bar{c}=0.5$, at 1/3, 2/3 and 1 of the stoppage time.}
	\label{fig:radial:2nd.eps}
\end{figure}

The last example reflects the effect of partial blockage of the injection hole. It can be seen that the upper half of the domain is mainly out of the flow domain.
\begin{figure}[!h]
	\includegraphics[width=\linewidth]{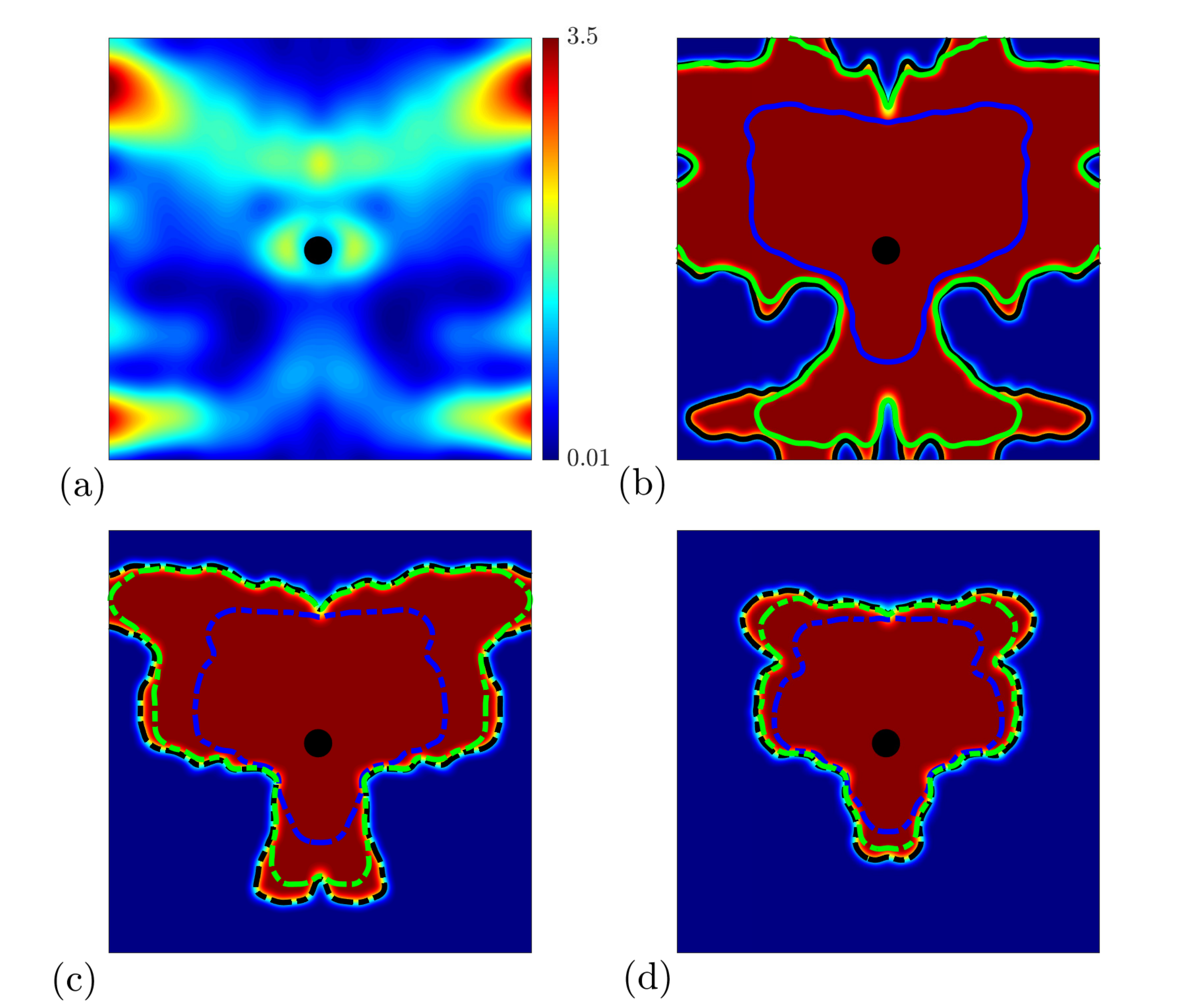}	
	\caption{The front evolution during displacement of a Newtonian fluid by a  Bingham fluid: (b) $Y=0.5$; (c) $Y=1$; (d) $Y=1.5$; all flows through the randomized geometry shown in panel (a). The blue, green, and black dashed lines show the displacing front, where $\bar{c}=1$, at 1/3, 2/3 and 1 of the stoppage time.}
	\label{fig:radial:3rd.eps}
\end{figure}

\begin{figure}[!h]
	\includegraphics[width=\linewidth]{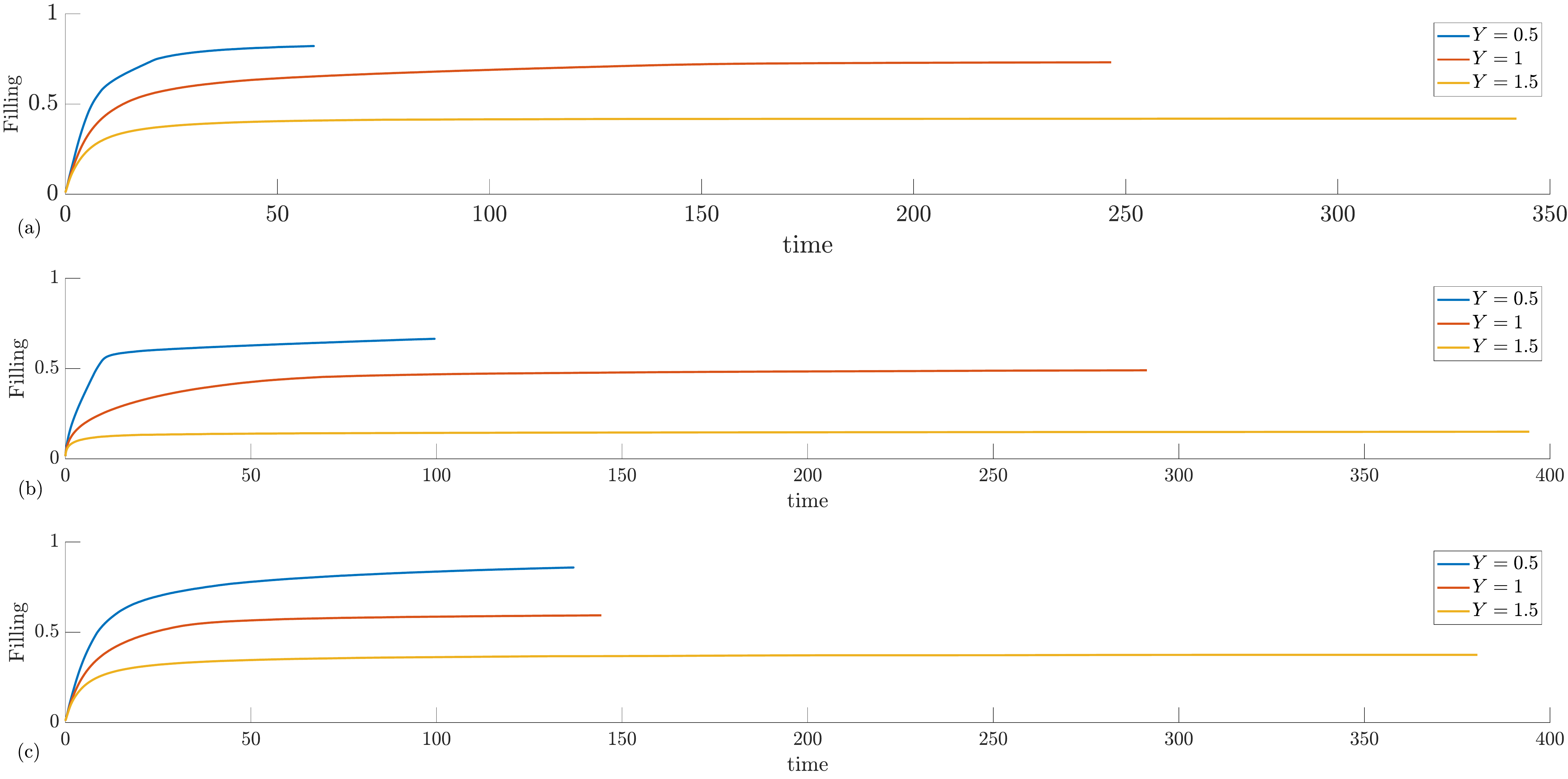}	
	\caption{The filling percentage by a Bingham fluid, shown by blue, red and yellow lines representing $Y=0.5$,-$Y=1$, and D-$Y=1.5$, respectively. Figures a), b), c) show the filling \% in the randomized geometries of Figs.~\ref{fig:radial:1st.eps}, \ref{fig:radial:2nd.eps} and \ref{fig:radial:3rd.eps}, respectively.}
	\label{fig:filling.eps}
\end{figure}

It can be seen that the initial inflow is fast: the pressure drop is concentrated mostly over a thin layer of invading fluid, driving it away from the perforation hole. However, later the invasion slows and is strongly influenced by the geometry. Figure \ref{fig:filling.eps} plots the evolution of the filling percentage, i.e.~the fraction of area occupied by $\bar{c}\geq 0.5$. We consistently see that lower $Y$ values approach their stoppage time faster and have higher filling percentage. Partly this is a viscous effect, in that the lower $Y$ value fluids move faster. It is not clear that the actual stoppage times have practical relevance.

\begin{figure}
	\centering
	\includegraphics[width=\linewidth]{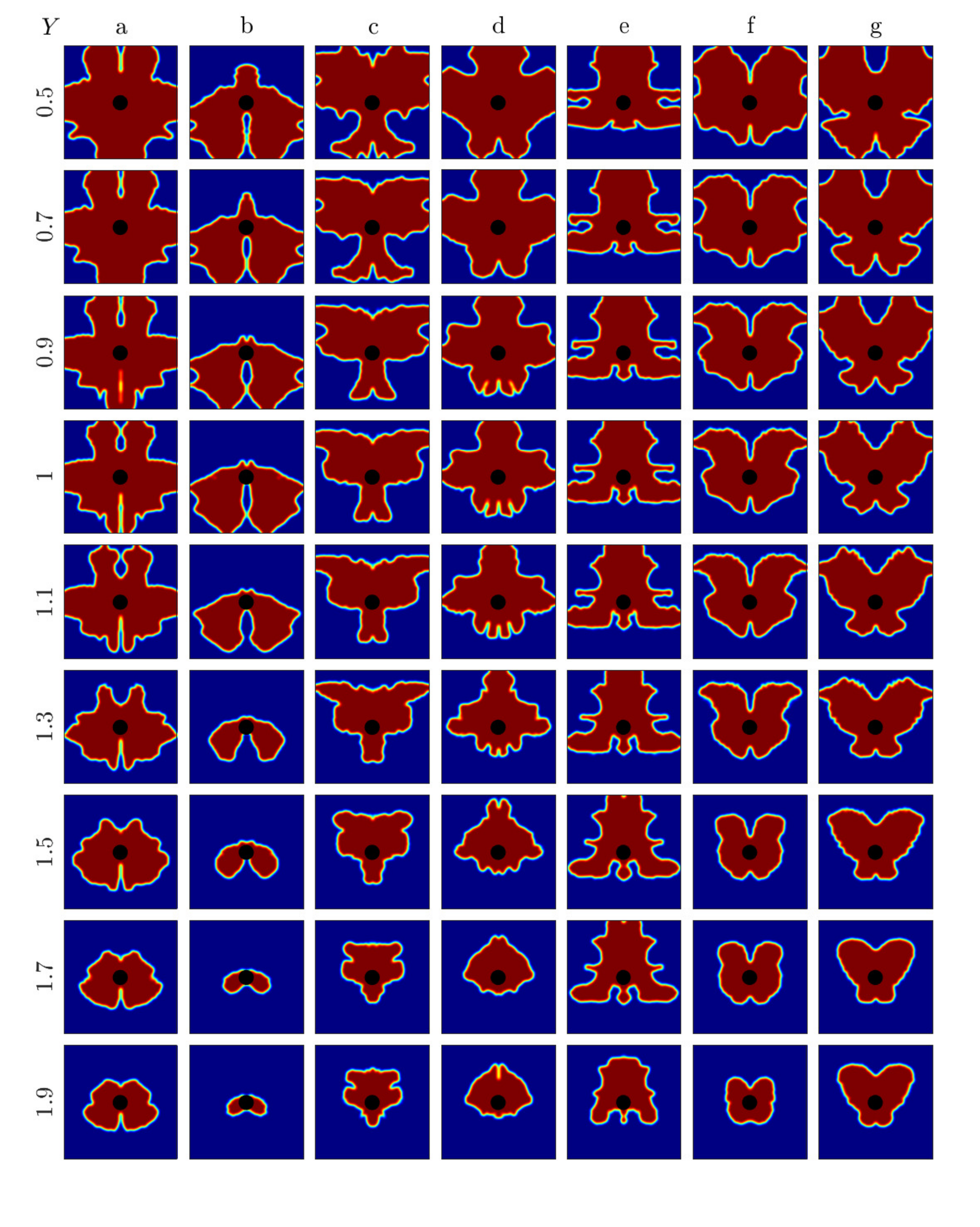}	
	\caption{Columns a, b and c  refer to invasion problems computed in the randomized geometries of Figs.~\ref{fig:radial:1st.eps}, \ref{fig:radial:2nd.eps} and \ref{fig:radial:3rd.eps}, respectively. Four other randomized geometries are considered in columns d-f. Each row denotes a different $Y$ value, as indicated. }
	\label{fig:final_all_cases}
\end{figure}

For the 3 microannulus geometries considered we have performed simulations over a wider range of $Y$ values. We present a panoramic view of the invasion at the final stoppage time in Fig.~\ref{fig:final_all_cases}, with columns a, b and c  referring the geometries of Figs.~\ref{fig:radial:1st.eps}, \ref{fig:radial:2nd.eps} and \ref{fig:radial:3rd.eps}, respectively. To these, we have added 4 further randomised geometries (columns d-f). The yield numbers for each row are given on the left hand side: $Y=0.5,0.7,0.9,...1.9$, with each column referring to a different geometry. Operationally, increasing $Y$ may be interpreted as either using a cement slurry of higher yield stress, or of applying a lower pressure drop.

Note that these geometries each have a microannulus gap thickness normalised to 1, so that each row is comparable. Operationally, one has no \emph{a priori} way of knowing the precise microannulus geometry and these geometries have been generated following the stochastic model of \cite{Trudel2023}, which is believed to reasonably represent the variation in microannulus thickness. Therefore, the significant differences seen in Fig.~\ref{fig:final_all_cases}, comparing between geometries at the same $Y$, reflect a significant effect of heterogeneity and \emph{luck}, in terms of the position of the perforation hole relative to the smallest gaps. For example, in geometry b for larger $Y$ it is clearly hard to penetrate much beyond the perforation vicinity, where Fig.~\ref{fig:radial:2nd.eps}a) shows that $H(x,y)$ is very small.

In terms of metrics that might be useful in representing the effectiveness of the squeeze job, firstly we make comparisons via 3 normalised parameters: $R_{n,min}$, $R_{n,max}$ and $R_{n,eq}$. $R_{n,min}$ and $R_{n,max}$ are the minimum and maximum values of the filling radius. For the filling radius we adopt a polar coordinate system $(r,\theta)$ fixed at the centre of the perforation hole. We take the length of the line connecting the filling front to the perforation hole, and divide by the length of the line (at the same $\theta$) that intersecting the border of the computational domain, i.e.~this length has a value $0.5\leq R\leq0.5\sqrt(2)$. The radius $R_{nd,eq}$ is the equivalent radius, computed as $\sqrt{\mbox{filled area}/\pi}$, divided by the maximum value $0.5/\sqrt{\pi}$.

\begin{figure}
	\includegraphics[width=\linewidth]{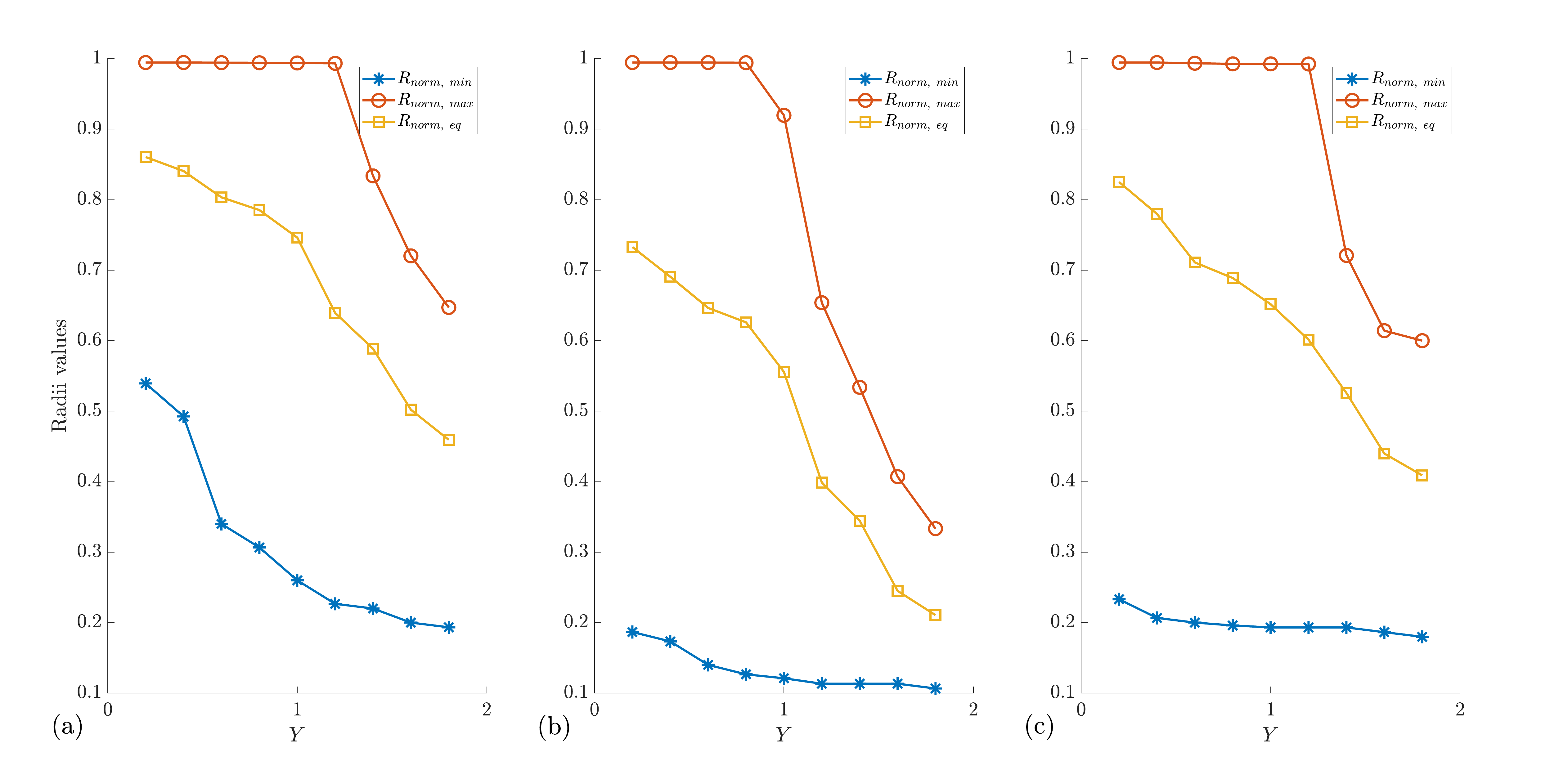}	
	\caption{Different invasion radii computed. Figures a), b) and c)  refer to invasion problems computed in the randomized geometries of Figs.~\ref{fig:radial:1st.eps}, \ref{fig:radial:2nd.eps} and \ref{fig:radial:3rd.eps}, respectively. The horizontal axis denotes $Y$.}
	\label{fig:radius_values}
\end{figure}

These radii are shown in Fig.~\ref{fig:radius_values}, plotted against $Y$ for the 3 sample geometries. The trends are intuitive: decreasing with $Y$ in all cases. In terms of practicality, with a view to the wide variability we have observed, one could envisage a probabilistic approach where e.g.~given assumed characteristics of the microannulus, one would like to say guarantee $R_{n,min} > 0.25$ (or dimensional equivalent) with confidence 95\%, for given fluid properties. The normalization used is mainly for comparison but also serves to map onto $1$ values where the invading fluid has reached the boundary of the computational domain. With reference to Fig.~\ref{fig:schematic2}, operationally there would be a stencil of perforation holes and the significance of $R_{n,max} \ge 1$ could be interpreted as a limit of when the invasion of one perforation meets that from another. To apply more practically, one would need to fit the computational domain to the perforation stencil. Again to apply this metric in a probabilistic way would require extensive sampling of microannulus geometries.

\begin{figure}[!h]
	\includegraphics[width=\linewidth]{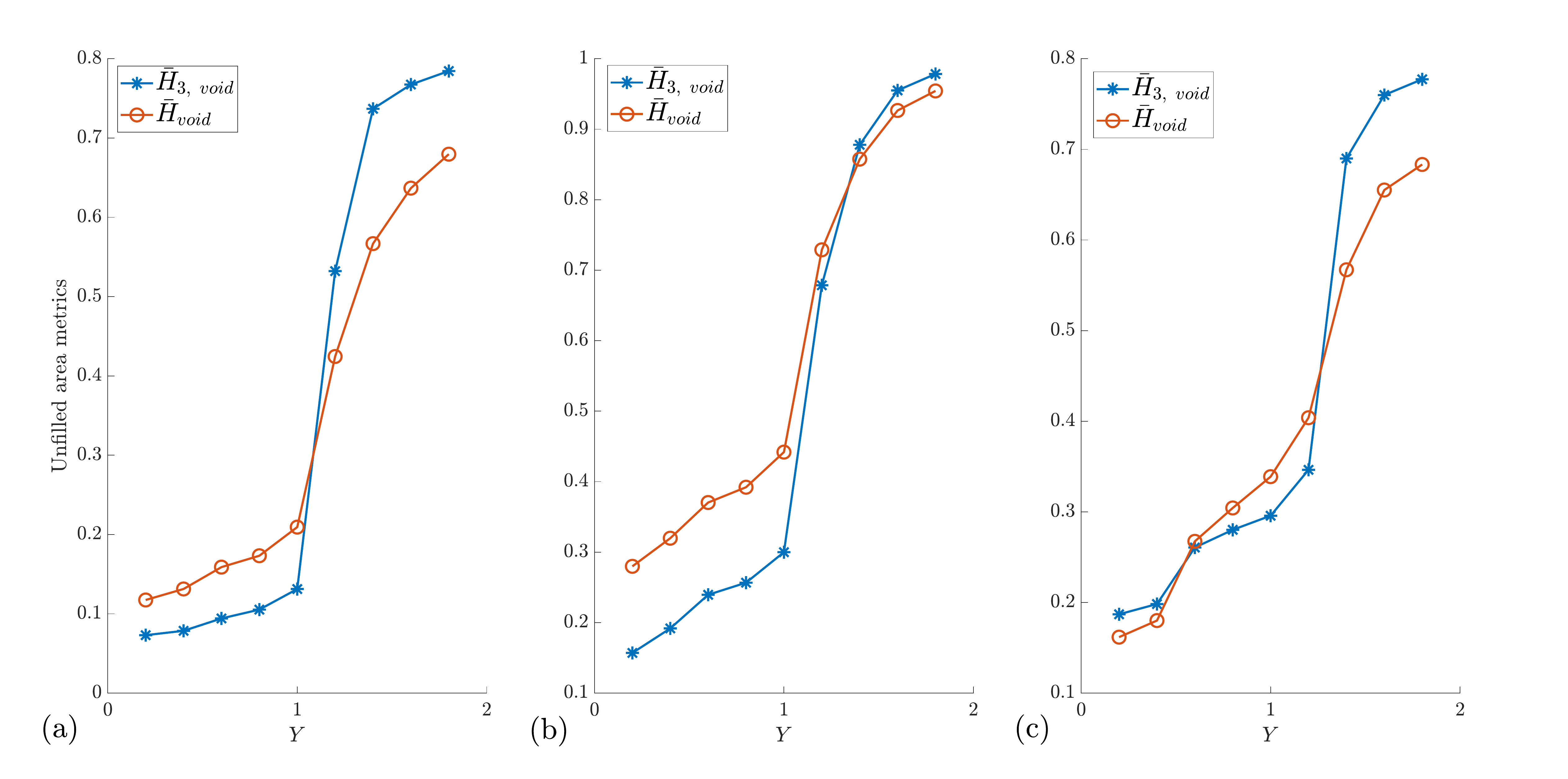}	
	\caption{Different measures of unfilled gap width computed. Figures a), b) and c)  refer to invasion problems computed in the randomized geometries of Figs.~\ref{fig:radial:1st.eps}, \ref{fig:radial:2nd.eps} and \ref{fig:radial:3rd.eps}, respectively. The horizontal axis denotes $Y$.}
	\label{fig:reducedGap}
\end{figure}

The variable  $R_{n,eq}$ clearly is a measure of the filled area. If we consider the overall objective of the squeeze cementing operation, to restore the well integrity, we might instead look at the volume fraction of the microannnulus that remains unfilled by cement:
\[ \bar{H}_{1,void} = \frac{\int_{\Omega} H(1-c)~dA}{\int_{\Omega} H~dA}   , \]
i.e.~this represents the volume remaining for the gas to flow through. A variation on this theme would be to compute $\bar{H}_{3,void}$:
 \[ \bar{H}_{3,void} = \frac{\int_{\Omega} H^3(1-c)~dA}{\int_{\Omega} H^3~dA}   . \]
The point of the cubic power is that the leakage flow rates are generally low and non-inertial, hence proportional to $H^3$. The metric $\bar{H}_{3,void}$ indicates the fraction of large leakage pathways remaining open, i.e.~we might think of interpreting $\bar{H}_{3,void}$ as an indicator of leakage reduction. These metrics are plotted in Fig.~\ref{fig:reducedGap}.

Lastly, one might also attempt to supplement the above calculated metrics with a simpler ``well integrity score'' (WIS). Again referring back to the perforation stencil of Fig.~\ref{fig:schematic2}, the main point of the squeeze operation might be to achieve integrity between adjacent perforation holes. Instead of computing a (very large) multi-perforation flow, one could interpret the single hole simulations in this context. For example, WIS $= 0$, if the edge of the domain is not reached; WIS $= k$, if the cement reaches to $k$ edges of the domain. Evidently, having large WIS would suggest that either the cement meets between the holes or at the least that the pathway that gas must follow through the perforated/squeezed section of well has got more tortuous and constrained. Table \ref{table:WIS} shows the WIS values computed from Fig.~\ref{fig:final_all_cases}.

\begin{table}[!h]
\begin{center}
 	\begin{tabular}{ |c|c|c|c| }
		
		\hline
		Y& a & b & c \\
			\hline
		0.5& 3 & 2 & 3 \\
		0.7& 3 & 2 & 1 \\
		0.9& 3 & 2 & 1 \\
		1.0& 3 & 2 & 1 \\
		1.1& 2 & 0 & 1 \\
		1.3& 1 & 0 & 1 \\
		1.5& 0 & 0 & 0 \\
		1.7& 0 & 0 & 0 \\
		1.9& 0 & 0 & 0 \\
		\hline
	\end{tabular}
\caption{WIS index for different $Y$ numbers (first column). Columns a, b and c  refer to invasion problems computed in the randomized geometries of Figs.~\ref{fig:radial:1st.eps}, \ref{fig:radial:2nd.eps} and \ref{fig:radial:3rd.eps}, respectively.}
\label{table:WIS}
\end{center}
\end{table}

\section{Summary}

In this paper we have introduced a type of invasion problem that has relevance to the remediation of oil and gas well leakage. At its heart, this involves the flow of a yield stress fluid along an uneven channel, typically displacing a less viscous preflush, i.e.~water. Apart from the remediation application, a similar process is undertaken for sealing around wells close to CCS reservoirs and there is also a wide variety of similar flows that occur in grouting. As long as the microannulus thickness varies slowly with distance, a Hele-Shaw type modelling approach is appropriate. It is perhaps ironic that the same approach is used for modelling the initial cementing flow, as is used here for repairing the defective cement seal. The flows are however simpler in that the positive viscosity ratio reduces dispersion and the smaller gap size renders buoyancy effects unimportant.

We have developed a displacement flow code that has been used to study 2 specific invasion geometries: planar and radial invasion. In the context of squeeze cementing, it
might be wondered where the planar invasion is relevant? With reference to Fig.~\ref{fig:schematic2}, the perforation gun pattern generally distributes the shots azimuthally but the axial spacing can be relatively tight (e.g.~4 shots per foot), although there must be intact metal between perforations. Therefore, having a long row of perforation holes in a helical arrangement is common. If the cement around each perforation is damaged and washed out, the result is a connected line of invasion holes, analogous to the planar front studied.

Nevertheless, our main focus has been on the radial invasion. Here we have taken a number of randomized microannulus geometries and studied the effects of the invasion parameters. In dimensionless terms, the key parameter is $Y$, denoting the effects of fluid yield stress divide by the imposed pressure drop. For each geometry one can say that the results are quite intuitive. As $Y$ increases the penetration is reduced and the repair (filling) of the microannulus is worse. Eventually, the cement slurry does not stop before it attains the boundary of the computational domain. Perhaps less intuitive is the degree of variability of the penetration behaviour with the specific microannulus geometry, captured visually well in Fig.~\ref{fig:final_all_cases}, i.e.~each microannulus here has a thickness normalised to $\bar{H}=1$ and each is constructed using the same sampling process from \cite{Trudel2023}. It is not hard to imagine that this same variability can account for the observed unreliability of the squeeze cementing operation.

We have post-processed our results to give practical and meaningful metrics that could be used as the basis of a probalistic risk-based design process, e.g.~being able to specify that the cement has penetrated a minimum distance $\hat{R}_{min}$ around each perforation hole, with confidence 95\%. Further post-processing has targeted metrics such as  $\bar{H}_{3,void}$ that are relevant to leakage flow rate. These are mainly examples, and a more refined analysis might try to either compute a full pattern of perforation holes (Fig.~\ref{fig:schematic2}), or to scale and orient the rectangular domain to represent a helical strip of microannulus, e.g.~with periodicity boundary conditions.

One concern with pursuing a risk-based procedure, as above, is that the computations we have carried out are expensive. The augmented Lagrangian method outlined and adopted is reliable in converging and in effectively resolving those parts of the annulus that are stationary, but is slow to converge. Here that procedure is repeated on each timestep. In a Monte-Carlo approach one would need many hundreds of microannulus geometries in order to generate suitable distributions. Therefore, a faster computational method is needed in order to develop this practically. One approach might be to directly calculate the final stopping distribution of the fluids, without recourse to the actual invasion transient.

\section*{Acknowledgements}
The authors gratefully acknowledge funding from the project: Plug and Abandon Strategies for Canada's Oil and Gas Wells, jointly supported by the AUPRF Program of PTAC (grant number 17-WARI-02) and by the Collaborative Research and Development Program of NSERC (grant number CRDPJ 516022–17). We also acknowledge the support from NSERC via scholarship number PGSD3 519200-2018, (ET).

\bibliography{Squeeze_model}

\end{document}